\documentclass[preprint,5p,times,twocolumn,authoryear]{elsarticle}
\usepackage{booktabs}
\usepackage{amsmath,amssymb,amsfonts}
\usepackage{graphicx}
\usepackage{textcomp}
\usepackage{float}
\usepackage{stfloats}
\usepackage{subfigure}
\usepackage{multirow}
\usepackage{makecell}
\usepackage{pifont}
\usepackage{xcolor}
\usepackage{threeparttable} 
\usepackage[ruled,linesnumbered]{algorithm2e}

\journal{Journal}

\begin{document}

\begin{frontmatter}

\title{AdvMIL: Adversarial Multiple Instance Learning for the Survival Analysis on Whole-Slide Images \tnoteref{title-lb1} \tnoteref{title-lb2}}

\tnotetext[title-lb1]{This work is under review.} 
\tnotetext[title-lb2]{Source code is available at https://github.com/liupei101/AdvMIL for facilitating further research.}

\author[afn1]{Pei Liu}
\ead{yuukilp@163.com}

\author[afn1]{Luping Ji\corref{cor1}}
\cortext[cor1]{Corresponding author: L. Ji (jiluping@uestc.edu.cn).}

\author[afn2]{Feng Ye}

\author[afn1]{Bo Fu}

\affiliation[afn1]{organization={School of Computer Science and Engineering, University of Electronic Science and Technology of China},
            addressline={Xiyuan Ave}, 
            city={Chengdu},
            postcode={611731}, 
            state={Sichuan},
            country={China}}
\affiliation[afn2]{organization={Institute of Clinical Pathology, West China Hospital, Sichuan University},
            addressline={Guo Xue Xiang}, 
            city={Chengdu},
            postcode={610041}, 
            state={Sichuan},
            country={China}}

\begin{abstract}
The survival analysis on histological whole-slide images (WSIs) is one of the most important means to estimate patient prognosis. Although many weakly-supervised deep learning models have been developed for gigapixel WSIs, their potential is generally restricted by classical survival analysis rules and fully-supervised learning requirements. As a result, these models provide patients only with a completely-certain point estimation of time-to-event, and they could only learn from the labeled WSI data currently at a small scale. To tackle these problems, we propose a novel adversarial multiple instance learning (AdvMIL) framework. This framework is based on adversarial time-to-event modeling, and integrates the multiple instance learning (MIL) that is much necessary for WSI representation learning. It is a plug-and-play one, so that most existing MIL-based end-to-end methods can be easily upgraded by applying this framework, gaining the improved abilities of survival distribution estimation and semi-supervised learning. Our extensive experiments show that AdvMIL not only could often bring performance improvement to mainstream WSI survival analysis methods at a relatively low computational cost, but also enables these 
methods to effectively utilize unlabeled data via semi-supervised learning. Moreover, it is observed that AdvMIL could help improving the robustness of models against patch occlusion and two representative image noises. The proposed AdvMIL framework could promote the research of survival analysis in computational pathology with its novel adversarial MIL paradigm. 
\end{abstract}

\end{frontmatter}

\section{Introduction}
Survival analysis, also known as time-to-event analysis, is one of the primary statistical approaches for analyzing data on time to event \citep{cox1975partial,kalbfleisch2011statistical}. It is usually adopted in medical fields to analyze clinical materials and assist doctors in understanding disease prognosis \citep{Wulczyn2021}. Histological whole-slide image (WSI) is one of these materials. It is produced by scanning tissue slides (millimeter scale) with a high-end microscope. Compared with other materials like demographics and genomics, digitized WSIs can present unique hierarchical 
views at a gigapixel-resolution \citep{zarella2018apractical}, \textit{e.g.}, tissue phenotype, tumor microenvironment, and cellular morphology. These rich and diverse microscopic information could provide valuable cues for the prognosis of tumor diseases \citep{Yu2016pred,CHEN2022865}, contributing to the improvement of patient management and disease outcomes \citep{Nir2018,Kather2019,Skrede2020}. 

\begin{figure}[ht]
\centering
\includegraphics[width=0.48\textwidth]{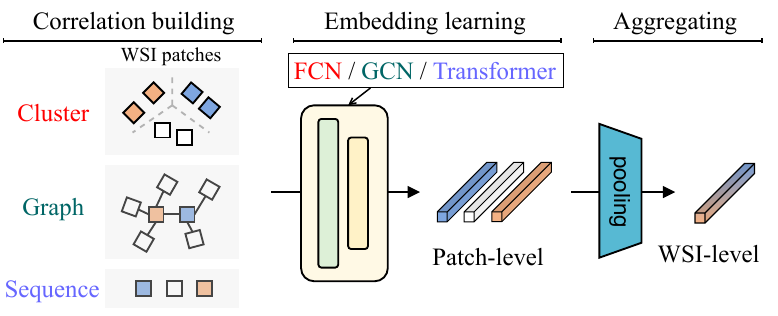}
\caption{Paradigm of embedding-level multiple-instance learning (MIL) for WSI representation learning. The structure of cluster, graph, or sequence is usually adopted to build patch correlations. These correlations are utilized by different networks to learn WSI-level representations.}
\label{fig:intro}
\end{figure}

\begin{figure*}[ht]
\centering
\includegraphics[width=\textwidth]{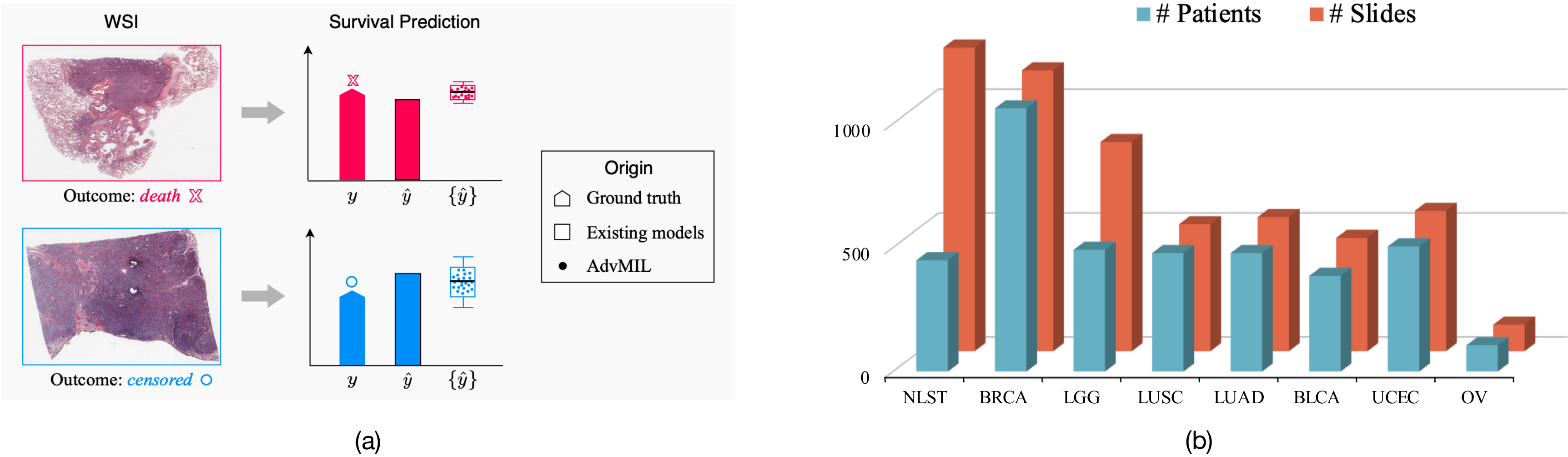}
\caption{The commonalities of existing WSI survival analysis models in terms of output and input:  (a) model output, existing methods are limited to a point time-to-event estimation, whereas ours can provide an estimation of time-to-event distribution, believed to be more robust and interpretable; (b) model input, all the most frequently-used datasets for WSI survival analysis are at a very small scale, usually with around 500 patients or 1,000 slides.}
\label{fig:intro-moti}
\end{figure*}

Unlike regular natural images, histological WSIs are usually with an extremely-high resolution, \textit{e.g.}, 40,000 $\times$ 40,000 pixels. This poses great challenges to WSI analysis and modeling, especially the global representation learning of WSIs. To tackle these challenges, many methods follow a weakly-supervised framework with three stages: i) WSI patching, ii) patch-level feature extracting, and iii) slide-level representation learning \citep{Chen2022scaling,GhaffariLaleh2022}. In procedure, this framework derives global representations through building patch correlations, learning patch-level embeddings, and aggregating patch-level embeddings, as shown in Figure \ref{fig:intro}. It is often cast as embedding-level multiple instance learning (MIL) \citep{ilse2018attn,carbonneau2018multiple}. According to the structure of patch correlation, the mainstream MIL methods for survival analysis can be roughly grouped into three categories: cluster-based \citep{yao2020whole,shao2021weakly}, graph-based \citep{li2018graph,chen2021whole}, and sequence-based \citep{huang2021integration,shen2022explainable,liu2022dual}. 

Although there are many kinds of methods for the survival analysis on gigapixel WSIs, these methods are generally restricted by the classical survival analysis rules that include a certain survival assumption regarding hazard function \citep{cox1975partial,kalbfleisch2011statistical,liu2021opt} and a likelihood estimation in discrete time domain \citep{zadeh2020bias}. This leads to a point survival estimation of them. However, the output of point estimation, compared with that of distribution estimation illustrated in Figure \ref{fig:intro-moti}(a), represents a single completely-certain result believed to be lacking in predictive robustness and interpretability \citep{lakshminarayanan2017simple,kendall2017uncertainties,nazarovs2022understanding,Linmans2023}. In addition, in terms of input, all these methods train their own networks with slide-level labels, following a fully-supervised learning setting. This means that their training needs sufficient labeled data to make the models well-generalized to unseen samples. However, among current publicly-available WSI datasets, patient number is often limited to around 500, as shown in Figure \ref{fig:intro-moti}(b). By contrast, the standard datasets in deep learning, such as ImageNet \citep{Deng2009}, often contains more than 10,000 samples. This fact indicates that current WSI-based survival analysis is still in a small data regime that is believed to have adverse effects on the generalization ability of deep learning models \citep{Chapelle2009,Zhou2021,Marini2021}. 

Generative adversarial network (GAN) \citep{goodfellow2014generative} offers means to mitigate these problems. On one hand, GAN is a generative model capable of estimating complex data distribution via implicitly sampling, naturally fit for predictive distribution modeling. On the other hand, the generator-discriminator structure of GAN can take fake (or unlabeled) samples as input, just meeting the needs of semi-supervised learning \citep{goodfellow2016nips}. In this way, GAN-based models are never limited to certain point estimation 
and fully-supervised learning; instead, they provide robust distribution estimations, and notably, they learn from unlabeled data to enhance their generalization ability \citep{springenberg2015unsupervised,salimans2016improved,miyato2018virtual,li2021triple}. 

In the past few years, GAN has attracted great attention and inspired many interesting applications beyond image generation \citep{gui2021areview}. Survival analysis is one of them. It is first combined with a conditional GAN (cGAN) \citep{mirza2014conditional}, referred to as adversarial time-to-event modeling, in order to develop a general assumption-free survival model for analyzing clinical tabular data \citep{chapfuwa2018adversarial,chapfuwa2020calibration}. Afterward, this general model is successfully extended to convolutional neural networks for the time-to-event analysis on CT images \citep{uemura2021weakly}. These models have shown promising results, especially their advantages of predictive accuracy and robustness. Further adoption of GAN for survival analysis in computational pathology is strongly anticipated, since the current survival analysis on WSI data still faces the problems posed by its conventional modeling paradigm aforementioned, \textit{i.e.}, classical survival analysis rules and fully-supervised learning. 

In this study, we propose a novel framework for the survival analysis on gigapixel whole-slide images, referred to as adversarial multiple instance learning (AdvMIL). This framework no longer relies on the classical paradigm of time-to-event modeling; instead, it is based on adversarial time-to-event modeling and integrates the multiple instance learning that is much necessary for WSI representation learning. As a result, most existing embedding-level MIL networks can be easily integrated into the proposed AdvMIL framework, thereby gaining the ability of survival distribution estimation and semi-supervised learning without introducing extra techniques. The results on three publicly-available WSI datasets verify that AdvMIL could often bring performance improvements to mainstream MIL networks at a relatively low computational cost; and most importantly, it could enable current models to effectively utilize unlabeled WSI data via semi-supervised learning. 

The primary contributions of this work are listed as follows. 

I) We present an adversarial multiple instance learning (AdvMIL) framework for the survival analysis on gigapixel whole-slide images. Most existing MIL-based WSI survival analysis methods can be upgraded by applying this framework, thereby gaining the abilities of survival distribution estimation and semi-supervised learning. To our best knowledge, the proposed AdvMIL is the first one to adopt GAN in computational pathology for survival analysis. 

II) We demonstrate how GAN can be combined with MIL paradigm to perform survival analysis on gigapixel WSIs, by the two key components of AdvMIL: the MIL encoder in generator and the fusion network with region-level instance projection (RLIP) in discriminator. 

III) We further explore how to train existing WSI models with our AdvMIL framework in a semi-supervised manner. Moreover, we propose a $k$-fold training strategy to make the semi-supervised learning more effective. 

IV) We validate the effectiveness of AdvMIL in predictive performance, semi-supervised learning, and model robustness, through the extensive experiments on a total of 3,101 WSIs from three publicly-available datasets. Empirical results suggest that AdvMIL could boost the development of survival analysis in computational pathology by its adversarial MIL paradigm. 

\section{Related work}

\subsection{Survival analysis of WSIs}
Predicting time-to-event from digitized gigapixel WSIs is an active research topic in recent years. Here we review some representative works, mainly focusing on their networks and survival loss functions. 

(1) \emph{Multiple-instance learning network} 

To learn global representations from gigapixel WSIs, MIL (see Figure \ref{fig:intro}) is widely adopted in end-to-end deep learning models \citep{GhaffariLaleh2022}. These models usually employ different backbones for different patch correlations, \textit{e.g.}, fully-connected networks for clusters, graph convolution networks \citep{kipf2016semi} for graphs, and Attention-MIL \citep{ilse2018attn} or Transformers \citep{Vaswani2017,dosovitskiy2020vit} for sequences. From an unified perspective, these networks first transfer patch features along patch correlations to learn non-local patch-level embeddings, and then extract WSI representation by pooling these embeddings into a global feature vector. 

(2) \emph{Survival loss function} 

Most models make some assumptions on hazard function \citep{kalbfleisch2011statistical} for survival analysis. Cox proportional hazard \citep{cox1975partial} and accelerated failure time \citep{wei1992aft} are the two most classical assumptions in them. The loss functions based on these two assumptions are very popular in WSI survival modeling, owing to their simplicity and interpretability. In addition, the survival loss function based on a maximum likelihood estimation, which has demonstrated good calibration and discrimination on tabular data \citep{zadeh2020bias}, is adopted by \cite{chen2021whole,Chen2022scaling} and \cite{liu2022dual} for WSI analysis. It requires a pre-discretization of survival time although it is assumption-free. These functions are derived from a class of \textit{discriminative} models, often leading to a limitation---the point estimation of time-to-event. 

\subsection{Adversarial time-to-event analysis}

(1) \emph{Generative adversarial network} 

GAN is a powerful \textit{generative} model \citep{goodfellow2014generative} with a wide variety of vision applications \citep{gui2021areview}, \textit{e.g.}, image generation, denoising, and editing. Additionally, it has also been extensively studied to perform semi-supervised learning \citep{springenberg2015unsupervised,carmon2019unlabeled} or enhance the robustness of deep learning networks \citep{goodfellow2014explaining,miyato2018virtual}. Following GAN, a conditional GAN (cGAN) \citep{mirza2014conditional} is first proposed to improve the quality of image generation by incorporating conditional labels. Based on cGAN, many efforts have been made to better exploit conditional labels, such as the projection discriminator \citep{miyato2018cgans}, originally developed for a general application of regular natural images. As a representative work of cGAN, the projection discriminator projects a conditional vector on input's final embedding. 

(2) \emph{Time-to-event modeling via conditional GAN} 

It is seen in DATE \citep{chapfuwa2018adversarial} for the first time, motivated by the fact that cGAN can capture complex data distribution \citep{goodfellow2014generative}. Like cGAN, DATE also uses a generator-discriminator architecture. Specifically, the generator, denoted as $G$, estimates a conditional distribution of time-to-event by 
\begin{equation}
    \hat{t}=G(x,\mathcal{N})\sim P_{t\vert x},
\end{equation}
where $x$ denotes a conditional input and $\mathcal{N}$ denotes a noise input. The discriminator, denoted as $D$, recognizes the real input $(x, t)$ or the fake input $(x, \hat{t})$, where $t$ is the label of time-to-event w.r.t. $x$. Such an adversarial generative way enables DATE to estimate the distribution of time-to-event via implicitly sampling from $G$, rather than via learning and optimizing the parameters of a priori distribution. 

DATE is a general assumption-free model of time-to-event analysis, originally applied to clinical tabular data. Following the framework of DATE, pix2surv \citep{uemura2021weakly} is further developed with convolutional networks for CT images. This research of time-to-event modeling on tabular data and CT images has benefited from the generative modeling paradigm, and has demonstrated better predictive accuracy and robustness. However, on pathological WSIs it still remains open. Moreover, how to perform semi-supervised learning and whether semi-supervised learning is effective for survival analysis, have not been studied yet in both DATE and pix2surv. 

In Section \ref{sec:method}, we show how the framework of adversarial time-to-event modeling can be generalized to MIL so as to perform survival prediction tasks on gigapixel WSIs. 

\begin{figure*}[tp]
\centering
\includegraphics[width=0.98\textwidth]{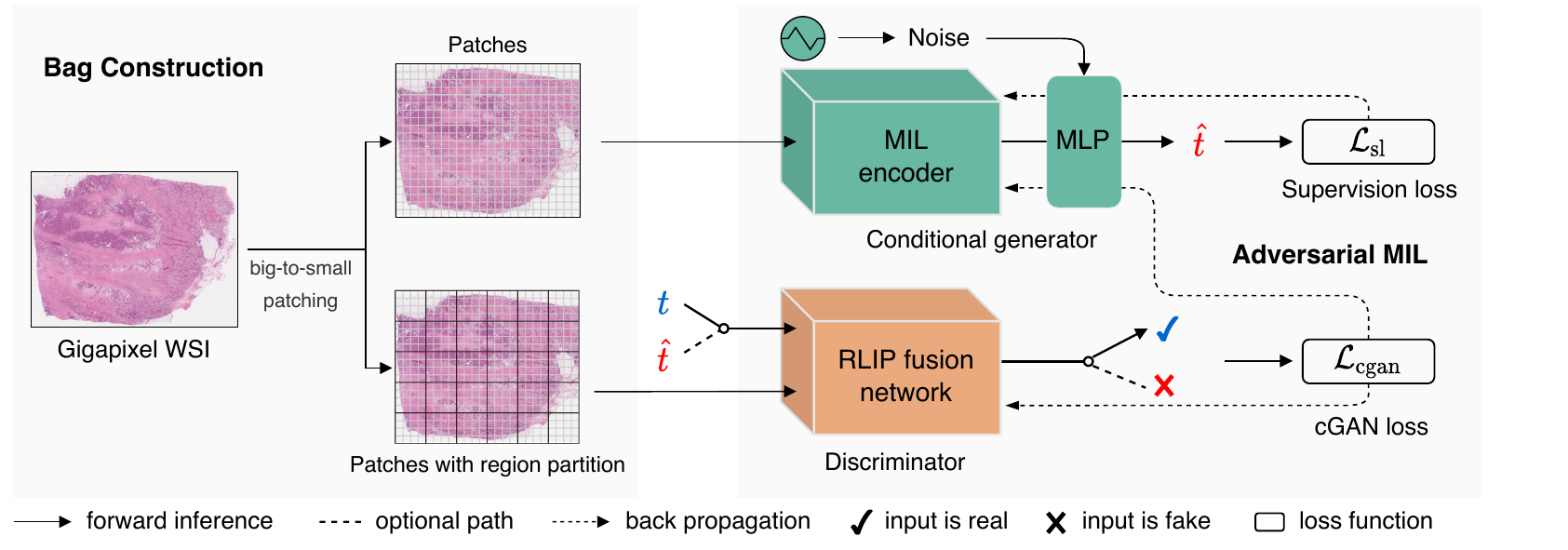}
\caption{AdvMIL overview. A gigapixel WSI is transformed into individual patches and the patches with equal region partition by big-to-small patching. A general MIL encoder is adopted as the backbone of conditional generator. An RLIP (region-level instance projection) fusion network is proposed to implement discriminator, elaborated in Figure \ref{fig0b}. Conditional generator is optimized by both $\mathcal{L}_{\mathrm{cgan}}$ and $\mathcal{L}_{\mathrm{sl}}$. Noise is given by a high-dimensional vector with certain distribution.}
\label{fig0a}
\end{figure*}

\section{Methodology}
\label{sec:method}
We show our Adversarial Multiple Instance Learning (AdvMIL) framework in Figure \ref{fig0a}. In a nutshell, AdvMIL generalizes adversarial time-to-event modeling to MIL by its two cores: the MIL encoder in generator and the fusion network with region-level instance projection (RLIP) in discriminator. At the end of this section, we describe the semi-supervised learning with AdvMIL, as well as a $k$-fold training strategy. 

\subsection{Preliminary}
\label{sec:prelim}
(1) \emph{Bag construction for WSIs}

As illustrated in Figure \ref{fig0a}, we prepare two types of patches in bag construction: i) the individual patches without any special structure for our generator, and ii) the patches with equal region partition for our discriminator. These two types are \textit{only} different in the way of organizing patches. 

Specifically, we design a big-to-small continuous-patching scheme. With this scheme, we can not only obtain the two types of patches described above, but also filter some patches in background to save computation cost. Namely, for one WSI, we first fix its magnification at $f\times$ ($20\times$ is a typical setting) and slice it into \textit{big} regions, each with the size of $\eta a \times \eta a$ pixels. In the meanwhile, some background regions are discarded. Then, we further slice each big region into $\eta^2$ \textit{small} patches (let $s=\eta^2$), each with the size of $a \times a$ pixels. Finally, we apply a feature extractor to all patches, leading to a bag of patch features. 

After the preprocessing steps given above, we denote the final bag of patch features from one patient by 
\begin{equation}
    X\in \mathbb{R}^{m\times c}: \{x_j\in \mathbb{R}^c\}_{j=1}^{j=m},
\end{equation}
where $m$ is the patch number in $X$, $c$ is the dimensionality of patch features, and $x_j$ denotes the $j$-th patch feature. On one hand, this given $X$ can be taken as individual patch features for our generator. On the other hand, it also can be viewed as the patch features with region partition for our discriminator, formulated as 
\begin{equation}
    X_r\in \mathbb{R}^{m\times c}: \{X_{\tau}\in \mathbb{R}^{s\times c}\}_{\tau=r_1}^{\tau=r_{m/s}},
\end{equation}
where $\{r_1, r_2, \cdots, r_{m/s}\}$ means the $m/s$ regions in $X$, $\tau$ is a region notation, and $X_{\tau}$ denotes the patch features of region $\tau$. Note that $m$ may be different across patients. Its common value is usually larger than 1,000 when $f=20$. 

(2) \emph{Notation convention} 

We denote survival data by $\mathcal{D}=\bigl\{(X_i,t_i,\delta_i)\bigr\}_{i=1}^{i=N}$, where $X_i, t_i \text{, and }\delta_i$ represent the bag features, follow-up time, and censorship status of the $i$-th patient, respectively. For the patients without censoring (\textit{i.e.}, with event occurrence), we denote their data by $\mathcal{D}_{e}=\bigl\{(X_i,t_i)\ \vert\ \delta_i=0 \text{ for } i=1,2,\cdots,N\bigr\}$. Similarly, the data of other patients with censoring (\textit{i.e.}, no event occurrence) are denoted as $\mathcal{D}_{ne}=\bigl\{(X_i,t_i)\ \vert\ \delta_i=1 \text{ for } i=1,2,\cdots,N\bigr\}$. For the $i$-th patient with $\delta_i=1$, we only know that its real time-to-event is strictly \textit{later} than $t_i$. 

\subsection{Adversarial multiple-instance learning}
Next, we formulate our AdvMIL framework in the form of multiple instance learning. 

(1) \emph{Generator}

To make our framework adapt to MIL and compatible with the existing networks for WSI survival analysis, we implement a new generator with a general MIL encoder and an MLP (multi-layer perceptron) layer. This encoder is adopted to extract global bag representations. It could be any embedding-level MIL networks that output bag-level feature vectors, \textit{e.g.}, cluster-, graph-, or sequence-based ones. The MLP layer is utilized to process both bag-level vectors and noise inputs, producing time-to-event estimations. 

For a given $X$ (the bag from one patient), we denote its time-to-event estimation by 
\begin{equation}
    \hat{t}=G(X,\mathcal{N}),\ \ \mathcal{N}\sim P_{n}=\mathbf{U}(0,1),
\end{equation}
where $\mathcal{N}$ is a variable that represents the noise with uniform distribution, and $G$ denotes a generator. In this way, $G$ implicitly estimates the distribution of time-to-event via sampling from $\mathcal{N}$. The distribution output from $G$ is expected to match the realistic conditional distribution of time-to-event on given $X$, \textit{i.e.}, $P_{t\vert X}$. By the noise-outsourcing lemma \citep{kallenberg2002foundations} from probability theory under minimal conditions, the existence of $G$ can be guaranteed \citep{zhou2022deep}, namely, 
\begin{equation}
\exists\ G, s.t.\  \hat{t}=G(X,\mathcal{N})\sim P_{t\vert X}.
\label{eq00}
\end{equation}
Since $\mathcal{N}$ is an independent variable, Equation (\ref{eq00}) can be further written into its joint distribution form, 
\begin{equation}
\bigl(X, G(X,\mathcal{N})\bigr)\sim P_{(X, t)}.
\label{eq01}
\end{equation}
$G$ can be determined by optimizing a cGAN \citep{mirza2014conditional} in which $X$ is taken as a conditional input to $G$. 

(2) \emph{Discriminator}

\begin{figure*}[tp]
\centering
\includegraphics[width=0.95\textwidth]{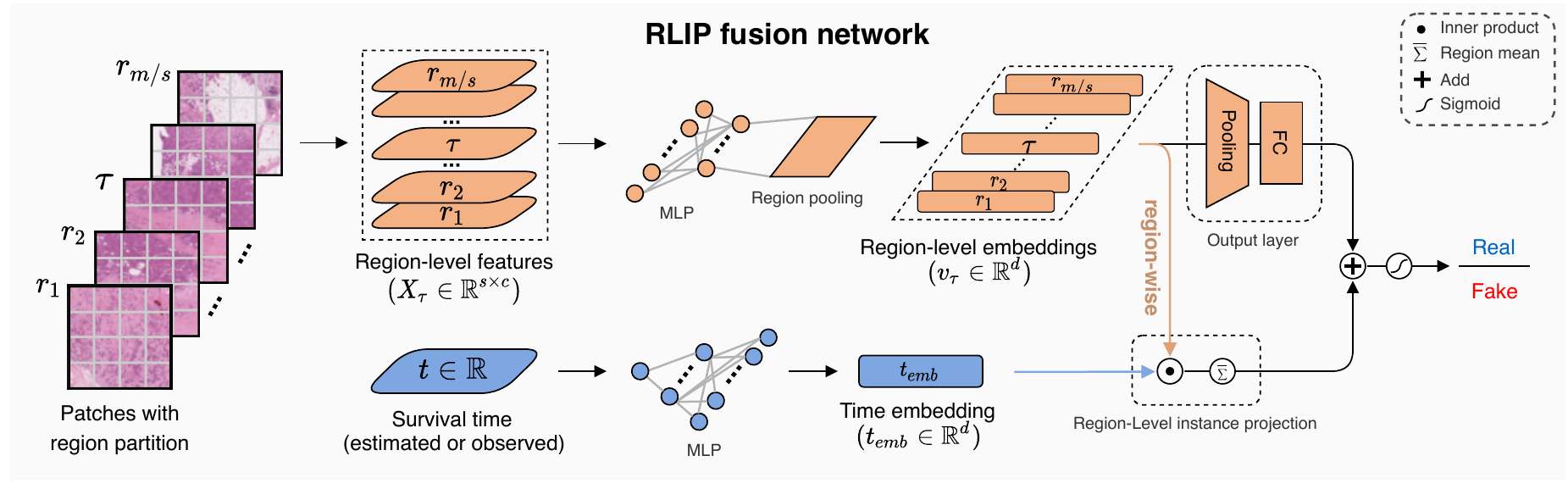}
\caption{Our fusion network with region-level instance projection (RLIP). As the discriminator of AdvMIL, it comprises two key parts: WSI region embedding and region-level instance projection. The patches with region partition are obtained by big-to-small patching (refer to Section \ref{sec:prelim} and Figure \ref{fig0a}). For any region $\tau\in\{r_1, r_2,\cdots, r_{m/s}\}$, $X_{\tau}$ represents its region-level features. $s$ indicates the patch number in each region. $v_{\tau}$ is the region-level embedding of $\tau$. MLP and FC mean multi-layer perceptron and fully-connected layer, respectively.}
\label{fig0b}
\end{figure*}

The discriminator, denoted as $D$, aims to distinguish between the fake pair $(X,\hat{t})$, sampled from $G$, and the real pair $(X, t)$, sampled from realistic data distribution. However, existing discriminators cannot be directly adopted to achieve this purpose, since $D$ is required to efficiently process the fusion of the given $X$ with an extremely-large matrix and the time $t$ with a single real value. To tackle this problem, we propose a novel fusion network with region-level instance projection (RLIP), as shown in Figure \ref{fig0b}. It can deal with the fusion of a big matrix and a scalar value, inspired by projection discriminator \citep{miyato2018cgans}. Moreover, our implementation of $D$ would not bring too much additional computation costs to existing mainstream MIL models, as shown in Section \ref{sec:comptoh}.

Our RLIP fusion network is illustrated in Figure \ref{fig0b}. After big-to-small patching (described in Section \ref{sec:prelim} and shown in Figure \ref{fig0a}), the patch features with region partition, $\{X_{\tau}\in \mathbb{R}^{s\times c}\}_{\tau=r_1}^{\tau=r_{m/s}}$, are pooled within each region to form the region-level embeddings $X_{emb}:\{v_{\tau}\in \mathbb{R}^{d}\}_{\tau=r_1}^{\tau=r_{m/s}}$. This is implemented by a region embedding layer $\phi: \mathbb{R}^{s\times c}\to \mathbb{R}^{d}$. Then, a time embedding, denoted as $t_{emb}\in\mathbb{R}^d$, is computed from survival time $t\in\mathbb{R}$ by an MLP $\varphi: \mathbb{R}\to \mathbb{R}^{d}$. This time embedding and the region-level embeddings are fused through a \textit{region-wise} inner product and a mean operation, represented by
\begin{equation}
    y_{fusion}=\frac{1}{m/s}\sum_{v_{\tau}\in X_{emb}}{v_\tau\cdot t_{emb}},
    \label{eq03}
\end{equation}
where $y_{fusion}\in\mathbb{R}$. Afterward, the region-level embeddings are fed into an output layer to yield a scalar output $y_{region}=\psi(\mathrm{gap}(X_{emb}))\in\mathbb{R}$, where $\mathrm{gap}(\cdot)$ and $\psi(\cdot)$ denote a global attention pooling function and a fully-connected layer, respectively. The final output of $D$ is $y_{D}=\mathrm{sigmoid}(y_{fusion}+y_{region}).$

(3) \emph{Network training}

As shown in Figure \ref{fig0a}, we use the two loss functions, cGAN loss and supervision loss, to optimize our network. cGAN loss is a general adversarial loss \citep{goodfellow2014generative} that involves $D$ and $G$, written as 
\begin{equation}
\begin{aligned}
\mathcal{L}_{\mathrm{cgan}} &= \mathop{\min}_{G}\mathop{\max}_{D}\mathbb{E}_{(X,t)\sim P_{\mathcal{D}_{e}}}\log D(X,t) \\
&+ \mathbb{E}_{X\sim P_X,\ \mathcal{N}\sim P_{n}}\log {\left[1-D(X,G(X,\mathcal{N}))\right]},
\end{aligned}
\label{eq1}
\end{equation}
where $D$ is optimized to accurately recognize real or fake pairs, while $G$ fools $D$ by generating the time-to-event that is expected to better match realistic data distribution. Note that $(X,t)$ is drawn from $P_{\mathcal{D}_{e}}$, instead of $P_{\mathcal{D}}$, in the first term of Equation (\ref{eq1}), because only the patients with event occurrence have the ground truth labels of time-to-event. 

Like most cGANs, the convergence of $\mathcal{L}_{\mathrm{cgan}}$ is guaranteed theoretically \citep{zhou2022deep}. Nevertheless, in network training, we observe that such an adversarial network is very difficult to be optimized when only using $\mathcal{L}_{\mathrm{cgan}}$. This problem is still open in adversarial learning \citep{goodfellow2016nips,gui2021areview}. To alleviate it, we additionally use an auxiliary supervision loss that fully utilizes time-to-event labels \citep{chapfuwa2018adversarial,uemura2021weakly}, to train the network and speed up its convergence. This auxiliary loss is defined by 
\begin{equation}
\begin{aligned}
\mathcal{L}_{\mathrm{sl}} = \mathop{\min}_{G}\mathbb{E}_{(X,t)\sim P_{\mathcal{D}_{e}},\ \mathcal{N}\sim P_{n}}\bigl\vert G(X,\mathcal{N})-t \bigr\vert\ +\\
 \mathbb{E}_{(X,t)\sim P_{\mathcal{D}_{ne}},\ \mathcal{N}\sim P_{n}}\text{ReLU}\bigl(t-G(X,\mathcal{N})\bigr),
\end{aligned}
\label{eq2}
\end{equation}
where $\mathcal{L}_{\mathrm{sl}}$ takes both censored and uncensored patients into account. Thus it could enable $G$ to narrow its empirical errors simultaneously on $P_{\mathcal{D}_{e}}$ and $P_{\mathcal{D}_{ne}}$. Unlike other loss functions that are widely adopted in WSI survival analysis, $\mathcal{L}_{\mathrm{sl}}$ no longer relies on any hazard function assumption or time pre-discretization, so our AdvMIL with $\mathcal{L}_{\mathrm{sl}}$ can estimate time-to-event distribution via implicitly sampling from $G$. 

The training procedure of AdvMIL, as shown in Algorithm \ref{apx:alg0}, is very similar to that of vanilla GANs. Specifically, it can also be viewed as an adversarial min-max game between $G$ and $D$. The key difference between AdvMIL and vanilla GANs lies in that AdvMIL training would involve censored patients ($i.e, \mathcal{D}_{ne}$) and auxiliary supervision loss ($i.e., \mathcal{L}_{\mathrm{sl}}$). 

\begin{algorithm}[tbp]
    \caption{Mini-batch training with AdvMIL (one epoch).}
    \label{apx:alg0}
    \KwIn{generator $G$, discriminator $D$, noise $\mathcal{N}$, batch size $b$,  cGAN loss $\mathcal{L}_{\mathrm{cgan}}$, supervision loss $\mathcal{L}_{\mathrm{sl}}$, and training data $\mathcal{D_{\mathrm{train}}}=\bigl\{(X_i,t_i,\delta_i)\bigr\}_{i=1}^{i=N_{\mathrm{train}}}$.}
    load mini-batch samples, $B \leftarrow \{(X_i,t_i,\delta_i)\}_{i=1}^{i=b}$
    
    \tcp{fix $G$, update $D$}
    
    $\mathcal{R}\leftarrow \{\}$ 
    
    \ForEach{$(X_i,t_i,\delta_i)$ in $B$}{
        \If{$\delta_i=0$}{
            $\hat{y}_{real}\leftarrow D(X_i,t_i)$ \tcp{real pairs}
            
            append $\hat{y}_{real}$ to $\mathcal{R}$
        }
        
        $\hat{y}_{fake}\leftarrow D(X_i,G(X_i,\mathcal{N}))$
        
        append $\hat{y}_{fake}$ to $\mathcal{R}$ \tcp{fake pairs}
        
    }
    
    update $D$ by optimizing the $\mathcal{L}_{\mathrm{cgan}}$ on $\mathcal{R}$ 
    
    \tcp{fix $D$, update $G$}
    
    $\mathcal{R}\leftarrow \{\}$, $\mathcal{R}_{sl}\leftarrow \{\}$ 
    
    \ForEach{$(X_i,t_i,\delta_i)$ in $B$}{
        $\hat{t}_i\leftarrow G(X_i,\mathcal{N})$
        
        append $(\hat{t}_i,t_i)$ to $\mathcal{R}_{sl}$
        
        $\hat{y}_{fake}\leftarrow D(X_i,\hat{t}_i)$ \tcp{fake pairs}
        
        append $\hat{y}_{fake}$ to $\mathcal{R}$
    }

    update $G$ by optimizing the $\mathcal{L}_{\mathrm{cgan}}$ on $\mathcal{R}$ and the $\mathcal{L}_{\mathrm{sl}}$ on $\mathcal{R}_{sl}$

\end{algorithm} 

\subsection{$k$-fold semi-supervised learning}
\label{sec:kssl}
Here we present how to perform semi-supervised learning with AdvMIL for WSI survival analysis, and then introduce our $k$-fold semi-supervised training strategy. 

We feed both labeled data (denoted as $\mathcal{D}_{\mathrm{l}}$) and unlabeled data (denoted as $\mathcal{D}_{\mathrm{ul}}$) into AdvMIL for semi-supervised learning. Specifically, for the unlabeled samples in $\mathcal{D}_{\mathrm{ul}}$, they will only be used by the second term of Equation (\ref{eq1}), \textit{i.e.}, $\mathop{\min}_{G}\mathop{\max}_{D}\mathbb{E}_{X\sim P_X,\ \mathcal{N}\sim P_{n}}\log {\left[1-D(X,G(X,\mathcal{N}))\right]}$, due to the absence of labels. Namely, unlabeled samples are utilized to infer time-to-event estimations through $G$, and then are further combined with these estimations to form fake pairs for optimizing the adversarial loss, as well as optimizing $G$ and $D$. 

Traditional semi-supervised learning usually directly utilizes all the unlabeled samples in $\mathcal{D}_{\mathrm{ul}}$ for mini-batch training. However, when the ratio of $\mathcal{D}_{\mathrm{ul}}$ is very high in training data, semi-supervised training would be dominated by those unlabeled samples, while often rarely focusing on the useful supervision signals from labeled samples. This problem could impair the efficiency of semi-supervised training, especially in a small data regime, \textit{e.g.}, WSI data; because a small data regime implies fewer labeled samples and weaker supervision signals. 

To decrease the influence of the problem above, we propose a new $k$-fold semi-supervised training strategy, shown in Algorithm \ref{apx:alg1}. It splits unlabeled data $\mathcal{D}_{\mathrm{ul}}$ into $k$ folds with the same size, $\mathcal{D}_{\mathrm{ul}}^{0}, \mathcal{D}_{\mathrm{ul}}^{1}, \ldots, \mathcal{D}_{\mathrm{ul}}^{k-1}$. At the $T$-th training epoch, only $\mathcal{D}_{\mathrm{ul}}^{\alpha}$ and labeled data $\mathcal{D}_{\mathrm{l}}$ are chosen as training samples, where $\alpha=(T\bmod k)$. We could see that if $k= 1$, this $k$-fold semi-supervised training strategy  exactly degenerates to traditional semi-supervised training.


\begin{algorithm}[tbp]
    \caption{$k$-fold semi-supervised training with AdvMIL.}
    \label{apx:alg1}
    \KwIn{labeled data $\mathcal{D}_{\mathrm{l}}=\bigl\{(X_i,t_i,\delta_i)\bigr\}_{i=1}^{i=N_{\mathrm{l}}}$, unlabeled data $\mathcal{D}_{\mathrm{ul}}=\bigl\{(X_i)\bigr\}_{i=1}^{i=N_{\mathrm{ul}}}$, the number of training epochs $\mathbb{T}$, the number of folds $k$.}
    
    
    randomly split $\mathcal{D_{\mathrm{ul}}}$ into $k$ folds: $\mathcal{D}_{\mathrm{ul}}^{0}, \mathcal{D}_{\mathrm{ul}}^{1}, \ldots, \mathcal{D}_{\mathrm{ul}}^{k-1}$
    
    \For{$T\leftarrow 0$ \KwTo $\mathbb{T}-1$}{
        $\mathcal{D_{\mathrm{train}}}\leftarrow \mathcal{D_{\mathrm{l}}} + \mathcal{D}_{\mathrm{ul}}^{T\bmod k}$
        
        mini-batch training on $\mathcal{D_{\mathrm{train}}}$ using Algorithm \ref{apx:alg0} 
        
        \tcp{Algorithm \ref{apx:alg0} will skip the steps involving $t_i$ or $\delta_i$ for unlabeled data}
    }

\end{algorithm} 

\section{Experiments and results}
\subsection{Experimental settings}

(1) \emph{Dataset description}

\begin{figure}[tbp]
\centering
\includegraphics[width=0.45\textwidth]{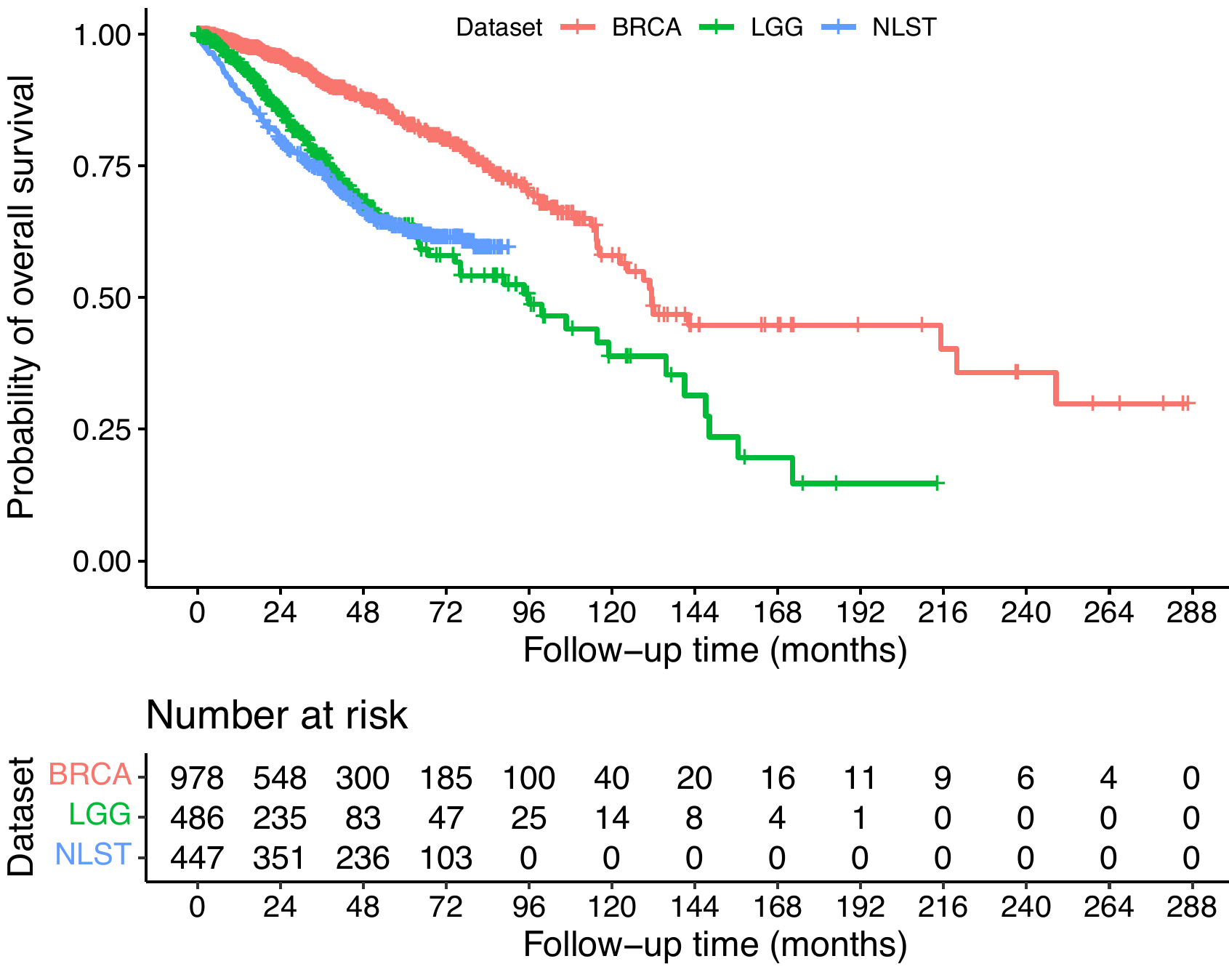}
\caption{Survival curves of patient cohorts. It is obvious that breast cancer patients have better overall survival and longer follow-up time.}
\label{fig:data-surv}
\end{figure}

There are three publicly-available WSI datasets used in this study. They are National Lung Screening Trial (NLST) \citep{nlst2011}, BReast CAncer (BRCA), and Low-Grade Glioma (LGG), where both BRCA and LGG are from The Cancer Genome Atlas (TCGA) \citep{Kandoth2013MutationalLA}. Our criterion of dataset selection mainly considers: i) the number of available patients, ii) whether having been adopted in previous publications, and iii) image quality. Finally, we collect a total of $3,101$ WSIs from $1,911$ patients without any subjective data curation. The overall survival curves of three datasets are 
shown in Figure \ref{fig:data-surv}, and dataset details are presented in Table \ref{apx:tb0}.

\begin{table}[htbp]
\small
\centering
\caption{Statistical details of three chosen WSI datasets.}
\label{apx:tb0}
\begin{tabular}{cccc}
\toprule
Items       & NLST & BRCA          & LGG          \\ \midrule
Death ratio          & 35.9\%      & 13.5\%        & 23.4\%       \\
\# Patients          & 447         & 978           & 486          \\
\# WSIs              & 1,222       & 1,043         & 836 \\    
\# Patches           & 3,955,344   & 3,228,480     & 2,637,456    \\
\# Patches / WSI & 3,236.8   & 3,095.4     & 3,154.9    \\ \bottomrule
\end{tabular}
\end{table}

(2) \emph{Chosen baselines}

\label{sec:baseline}
The four methods, ABMIL \citep{ilse2018attn}, DeepAttnMISL \citep{yao2020whole}, PatchGCN \citep{chen2021whole}, and ESAT \citep{shen2022explainable}, are chosen as our baselines, because i) they are derived from three mainstream categories ($i.e.$, cluster, graph, and sequence) for end-to-end WSI modeling and ii) they are the most representative ones of each category. We denote a baseline model by $\mathcal{M}$, and it could be one of ABMIL, DeepAttnMISL, PatchGCN, and ESAT. To validate the effectiveness of AdvMIL on four baselines, we compare three different models, as follows:
\begin{itemize}
    \item $\mathcal{M}_{\text{origin}}$, which follows the implementation of $\mathcal{M}$, with original MIL encoder and survival loss function. 
    \item $\mathcal{M}_{\text{base}}$,  which employs the original MIL encoder of $\mathcal{M}$, with the new $\mathcal{L}_{\mathrm{sl}}$ defined by Equation (\ref{eq2}).
    \item $\mathcal{M}$ + AdvMIL, which adopts a new AdvMIL scheme of a discriminator and a generator with the original MIL encoder of $\mathcal{M}$, and specially utilizes two new $\mathcal{L}_{\mathrm{sl}}$ and $\mathcal{L}_{\mathrm{cgan}}$ in adversarial learning.
\end{itemize}

(3) \emph{Implementation details}
\label{sec:impdel}

\textbf{Bag construction.} 
The magnification of each WSI is set to $20\times$ ($i.e., f=20$). For big-to-small patching, we set the patch number in each region to 16 ($s=\eta^2=16$), and patch size to $256\times 256$ pixels ($i.e., a=256$), so as to well balance computation efficiency and microscopic details. In addition, following CLAM \citep{lu2021data}, the feature extractor (with $c=1024$) applied to patches is the official truncated ResNet-50 model \citep{He2016}, pre-trained on the ImageNet \citep{Deng2009}. 

\textbf{Generator.} We empirically set the output dimensionality of MIL encoder to 384 for both sequence-based models and cluster-based ones, and 128 for graph-based ones (as it is efficient to train dense and large graphs). The MLP at the end of generator $G$ only contains two layers. In each layer, when random noise $\mathcal{N}$ is added, its dimentionality is set to the same as input's. For simplicity, we use a binary code to represent whether random noise is added to a layer of MLP. Codes 0 and 1 indicate not adding and adding random noise to the corresponding layer of MLP, respectively. Thus, for adversarial learning, all the possible settings of noise adding are 0-1, 1-0, and 1-1 in our two-layer MLP. For example, 1-0 indicates that noise is added to the first layer but not to the second one, and so on. 

\begin{table}[tp]
\small
\centering
\caption{Full training hyper-parameter settings.}
\label{apx:tb1}
\begin{tabular}{cc}
\toprule
Hyper-parameters      & Values          \\ \midrule
epoch number & 300 \\
batch size & 1 \\
gradient accumulation step & 16 \\
early-stopping patience & 30 \\
early-stopping warm-up & 5 \\
learning rate for $G$ & 0.00008 \\
learning rate for $D$ & 0.00008 \\
optimizer & adam \\
weight decay rate & 0.0005 \\
\bottomrule
\end{tabular}
\end{table}

\textbf{Discriminator.} For the RLIP fusion network $D$, we use an MLP and an average pooling layer to implement the region embedding layer $\phi$. This $\phi$ can transform the region features with any number of patches into a single region-level feature vector, \textit{i.e.}, $\phi$ is capable of working under various settings of patch regions (even if these regions have different patch numbers). This property indicates that region partition can be flexible 
and scalable. The feature dimensionality of $X_{emb}$ and $t_{emb}$ is $d=128$. The time embedding layer $\varphi$, which implements $\mathbb{R}\to \mathbb{R}^{128}$, is an MLP with two layers. The fully-connected layer $\psi$ implements $\mathbb{R}^{128}\to \mathbb{R}$.

\textbf{Network training.} The generator $G$ is updated by a simplified version of the second term of $\mathcal{L}_{\mathrm{cgan}}$, as follows: 
\begin{equation}
\mathcal{L}_{\mathrm{cgan}}(G)=\mathop{\max}_{G}\mathbb{E}_{X\sim P_X,\ \mathcal{N}\sim P_{n}} D\big(X,G(X,\mathcal{N})\big).
\end{equation}
It often helps to train adversarial networks \citep{salimans2016improved}. Our full training hyper-parameter settings are shown in Table \ref{apx:tb1}. The learning rate of $G$ decays by a factor of 0.5 if validation loss does not decrease within 10 epochs. Moreover, all these hyper-parameters always keep unchanged on three used WSI datasets. More experimental details can be seen in our publicly-available source codes. 

(4) \emph{Evaluation metrics}

We report the metrics frequently-used in survival analysis \citep{harrell1984regression}, Concordance Index (C-Index). It measures the model ability of risk discrimination or ranking, just like the AUC in classification evaluation. We also evaluate the Mean Absolute Error (MAE) of estimated $\hat{t}$ on ground truth labels, as calculated in Equation (\ref{eq2}). During evaluation, for any patient we get its $\hat{t}$ by randomly sampling from generator for 30 times (as it is relatively time-consuming to infer once on gigapixel WSIs). Given limited sampling times, we use the median of $\hat{t}$ as the final estimation for performance evaluation, because it is often more robust to outliers than a mean. 

Moreover, we use 5-fold Cross-Validation (CV) to evaluate each model. In training, the 20\% samples of training set are randomly picked out to form a validation set for early stopping and model selection. Our data splitting is conducted at patient-level. All experiments run on a workstation with 2 $\times$ NVIDIA V100s (32G) GPU. 

\subsection{Results and analysis}
(1) \emph{Overall performance}
\label{sec:overperf}

\begin{table*}[tp]
\small
\centering
\caption{Performance comparisons of different models.
The binary code in bracket represents one specific setting of the random noise added into the two-layer MLP designed at the end of $G$. Codes 0 and 1 indicate not adding and adding random noises to the corresponding layer of MLP, respectively. Bold number indicates the best metric on a given dataset, and subscript number is a standard deviation of performance metrics.}
\begin{threeparttable}[b]
\label{tb1}
\begin{tabular}{p{0.5cm}<{\centering}cp{1.4cm}<{\centering}p{1.4cm}<{\centering}p{1.4cm}<{\centering}ccc}
\toprule
\multicolumn{2}{c}{\multirow{2}{*}{Model}} & \multicolumn{3}{c}{C-Index $\uparrow$}             & \multicolumn{3}{c}{MAE $\downarrow$}                 \\ \cmidrule(lr){3-5} \cmidrule(lr){6-8}
                            &       & NLST        & BRCA        & LGG         & NLST        & BRCA        & LGG         \\ \midrule
\multirow{9}*{\rotatebox{90}{Sequence-based}} & ESAT$_{\text{origin}}$ \tnote{\dag}     & $0.661_{0.039}$ & $0.544_{0.034}$ & $0.448_{0.037}$ & $0.2054_{0.0263}$ & $0.0529_{0.0108}$ & $0.0663_{0.0087}$ \\
& ESAT$_{\text{base}}$ \tnote{\ddag}    & $0.653_{0.047}$ & $0.542_{0.063}$ & $0.638_{0.091}$ & $0.1920_{0.0261}$ & $\textbf{0.0344}_{0.0056}$ & $\textbf{0.0518}_{0.0060}$ \\
& ESAT + AdvMIL (0-1)                   & $\textbf{0.672}_{0.048}$ & $0.545_{0.065}$ & $0.621_{0.063}$ & $0.1871_{0.0203}$ & $0.0383_{0.0081}$ & $0.0526_{0.0058}$ \\
& ESAT + AdvMIL (1-0)                   & $0.649_{0.039}$ & $\textbf{0.562}_{0.067}$ & $\textbf{0.642}_{0.076}$ & $0.1995_{0.0240}$ & $0.0349_{0.0065}$ & $0.0522_{0.0049}$ \\ 
& ESAT + AdvMIL (1-1)                   & $0.660_{0.042}$ & $0.545_{0.075}$ & $0.634_{0.086}$ & $\textbf{0.1849}_{0.0153}$ & $0.0366_{0.0065}$ & $0.0523_{0.0051}$ \\ \cmidrule(lr){2-8}
& ABMIL$_{\text{base}}$ \tnote{\ddag}    & $0.525_{0.095}$ & $0.502_{0.051}$ & $0.493_{0.046}$ & $0.2280_{0.0333}$ & $\textbf{0.0387}_{0.0051}$ & $0.0651_{0.0048}$ \\
& ABMIL + AdvMIL (0-1)                   & $0.522_{0.069}$ & $\textbf{0.566}_{0.035}$ & $0.494_{0.023}$ & $0.2285_{0.0375}$ & $0.0433_{0.0070}$ & $0.0692_{0.0057}$ \\
& ABMIL + AdvMIL (1-0)                   & $0.531_{0.090}$ & $0.523_{0.057}$ & $\textbf{0.535}_{0.029}$ & $\textbf{0.2261}_{0.0361}$ & $0.0397_{0.0060}$ & $\textbf{0.0596}_{0.0095}$ \\ 
& ABMIL + AdvMIL (1-1)                   & $\textbf{0.544}_{0.084}$ & $0.550_{0.073}$ & $0.511_{0.065}$ & $0.2277_{0.0390}$ & $0.0419_{0.0059}$ & $0.0655_{0.0049}$  \\ \cmidrule(lr){1-8}
\multirow{5}*{\rotatebox{90}{Cluster-based}} & DeepAttnMISL$_{\text{origin}}$ \tnote{\dag}     & $\textbf{0.548}_{0.081}$ & $0.496_{0.048}$ & $\textbf{0.593}_{0.062}$ & - & - & - \\
& DeepAttnMISL$_{\text{base}}$ \tnote{\ddag}    & $0.543_{0.072}$ & $0.508_{0.049}$ & $0.517_{0.080}$ & $0.2262_{0.0343}$ & $\textbf{0.0398}_{0.0060}$ & $0.0703_{0.0069}$ \\
& DeepAttnMISL + AdvMIL (0-1)                   & $0.528_{0.051}$ & $0.498_{0.036}$ & $0.452_{0.072}$ & $0.2456_{0.0356}$ & $0.0450_{0.0066}$ & $0.0707_{0.0069}$ \\
& DeepAttnMISL + AdvMIL (1-0)                   & $0.531_{0.068}$ & $0.510_{0.075}$ & $0.510_{0.062}$ & $0.2260_{0.0334}$ & $0.0414_{0.0055}$ & $\textbf{0.0640}_{0.0028}$ \\ 
& DeepAttnMISL + AdvMIL (1-1)                   & $0.538_{0.048}$ & $\textbf{0.552}_{0.029}$ & $0.506_{0.054}$ & $\textbf{0.2258}_{0.0325}$ & $0.0421_{0.0053}$ & $0.0664_{0.0042}$ \\ \cmidrule(lr){1-8}
\multirow{5}*{\rotatebox{90}{Graph-based}} & PatchGCN$_{\text{origin}}$ \tnote{\dag}     & $0.605_{0.024}$ & $0.542_{0.072}$ & $0.585_{0.055}$ & -  & - & - \\
& PatchGCN$_{\text{base}}$ \tnote{\ddag}    & $0.579_{0.080}$ & $0.518_{0.049}$ & $0.592_{0.071}$ & $0.2131_{0.0375}$ & $\textbf{0.0405}_{0.0046}$ & $0.0634_{0.0052}$ \\
& PatchGCN + AdvMIL (0-1)                   & $0.613_{0.047}$ & $\textbf{0.562}_{0.094}$ & $0.560_{0.073}$ & $0.1944_{0.0266}$ & $0.0408_{0.0040}$ & $0.0590_{0.0056}$ \\
& PatchGCN + AdvMIL (1-0)                   & $0.580_{0.060}$ & $0.556_{0.063}$ & $0.561_{0.101}$ & $0.2080_{0.0327}$ & $0.0478_{0.0081}$ & $0.0561_{0.0054}$ \\ 
& PatchGCN + AdvMIL (1-1)                   & $\textbf{0.644}_{0.050}$ & $0.535_{0.091}$ & $\textbf{0.593}_{0.065}$ & $\textbf{0.1895}_{0.0175}$ & $0.0427_{0.0085}$ & $\textbf{0.0529}_{0.0043}$ \\
 \bottomrule
\end{tabular}
  \begin{tablenotes}
       \item [\dag] \scriptsize We adopt their original survival loss functions to report these results. The ABMIL with its original survival loss function is not reported, because ABMIL is originally proposed for classification. MAE is not reported for original DeepAttnMISL and PatchGCN since they do not provide the prediction of continuous survival time.
        \item [\ddag] \scriptsize We adopt the survival loss function given by Equation (\ref{eq2}) to report these results.
  \end{tablenotes}
\end{threeparttable}
\end{table*}

\textbf{The sequence-based ESAT with AdvMIL.} We first apply AdvMIL to sequence-based ESAT and test its performance, as ESAT is built upon the Transformer \citep{Vaswani2017} that has demonstrated remarkable success in various applications. Compared with ESAT$_{\text{base}}$, the model combining ESAT and AdvMIL is additionally optimized by the adversarial loss $\mathcal{L}_{\mathrm{cgan}}$. 

As shown in Table \ref{tb1}, we can see that 1) ESAT + AdvMIL could often obtain better performances than original ESAT in both C-Index and MAE, and its improvement on LGG is evident; 2) the model combining ESAT and AdvMIL outperforms its counterpart ESAT$_{\text{base}}$, by a C-Index improvement of 2.91\%, 3.69\%, and 0.63\% on NLST, BRCA, and LGG, respectively; 3) Applying AdvMIL to ESAT decreases MAE by 3.70\% on NLST, and slightly increases MAE by around +1\% on the other two datasets. These observations suggest that the proposed AdvMIL has competitive advantages over sequence-based ESAT, especially in terms of C-Index.  Intuitively, by adversarial learning $\mathcal{L}_{\mathrm{cgan}}$ could help to optimize $G$, and then make $G$ become more robust to the input space extended and smoothed by noise $\mathcal{N}$ \citep{miyato2018virtual}. In addition, it is observed that noise setting also affects model performance. This implies that a proper noise setting (the focus in adversarial training \citep{goodfellow2014explaining}) may help achieving better performance. More discussions on this can be seen in Section \ref{sec:ablation}. 

\textbf{Other mainstream MIL networks with AdvMIL.} We test the adaptability of AdvMIL to other mainstream MIL networks (with distinct backbones), and experimental results are shown in Table \ref{tb1}. From these results, we can see that 1) ABMIL + AdvMIL almost always outperforms its counterpart ABMIL$_{\text{base}}$ in terms of both C-Index and MAE, except the MAE on BRCA; 2) DeepAttnMISL + AdvMIL is obviously better in C-Index than its two baselines on BRCA, while performs worse on NLST and LGG; 3) After applying AdvMIL to PatchGCN, the combined model is comparable or even better than PatchGCN$_{\text{origin}}$ and PatchGCN$_{\text{base}}$ in most cases. 

These empirical results from Table \ref{tb1} suggest that AdvMIL could often help boosting the performance of mainstream MIL networks. Moreover, by comparing the results of four representative MIL networks, we notice that ESAT + AdvMIL achieves the best overall performance across three datasets.

It is worth noting that MAE is more sensitive to the value of $\hat{t}$ than C-Index since C-Index measures the quality of estimation ranking---it only considers relative errors in measurement. Given the fact that limited estimations are drawn from $G$ for evaluation, the final estimation of $\hat{t}$ may be more likely to stray from the expectation of distribution, \textit{i.e.}, $\mathbb{E}_{\mathcal{N}\sim P_{n}}P(t\vert X,\mathcal{N})$, resulting in worse MAEs, as shown in Table \ref{tb1}. 

(2) \emph{Computational overhead analysis}
\label{sec:comptoh}

We evaluate the computation efficiency of discriminator in model size and inference cost, to verify if the proposed discriminator will bring substantial computational overheads to mainstream MIL networks. Specifically, we measure the number of trainable parameters (\# Params) for all used models. Moreover, we assess the theoretical amount of Multiply–Accumulate operations (\# MACs) in models, by randomly selecting a patient from NLST to infer the model once. This patient has a total of 3,360 patches in its WSI bag. A Python toolkit, \textit{ptflops}, is adopted to calculate \# Params and \# MACs.

\begin{table}[tp]
\small
\centering
\caption{Model efficiency in terms of \# Params and \# MACs.}
\label{apx:tb2}
\begin{tabular}{cccc}
\toprule
\multicolumn{2}{c}{Network}       & \# Params & \# MACs \\ \midrule
\multicolumn{2}{c}{$D$ (ours)}              &  \textbf{206.34K}  & \textbf{452.08M}      \\ \cmidrule(lr){1-2}
\multirow{4}{*}{$G$} & ABMIL &  985.54K  & 2.31G \\
                   & DeepAttnMISL &  985.54K  & 1.33G \\
                   & PatchGCN   & 280.51K & 662.77M  \\
                   & ESAT   & 1.73M & 1.61G \\ \bottomrule
\end{tabular}
\end{table}

As shown in Table \ref{apx:tb2}, our discriminator only increases extra 206.34K trainable parameters and 452.08M MACs, obviously smaller than four generators. These results imply that it is acceptable to introduce such a discriminator into existing mainstream MIL networks, in terms of computational overhead. 

(3) \emph{$k$-fold semi-supervised training}

We further investigate the potential of adopting AdvMIL in semi-supervised learning. This is not studied in both of the two classical works of adversarial survival analysis, \textit{i.e.}, DATE \citep{chapfuwa2018adversarial} and pix2surv \citep{uemura2021weakly}. For simplicity, we denote training set and test set by $\mathcal{D}_{\mathrm{train}}$ and $\mathcal{D}_{\mathrm{test}}$, respectively. To conduct this experiment, we prepare unlabeled data at first. Specifically, we randomly select some samples from $\mathcal{D}_{\mathrm{train}}$ and mask their labels. These selected samples are label-unavailable in training, called unlabeled data and denoted as $\mathcal{D}_{\mathrm{ul}}$. The remaining samples are labeled data, denoted as $\mathcal{D}_{\mathrm{l}}=\mathcal{D}_{\mathrm{train}}\setminus \mathcal{D}_{\mathrm{ul}}$. Their labels are available in training. We set the ratio of labeled data $\frac{\mathcal{D}_{\mathrm{l}}}{\mathcal{D}_{\mathrm{train}}}$ to 0.2, 0.4, 0.6, 0.8 and 0.9. Then, we test and compare semi-supervised AdvMIL models and fully-supervised ones. (I) On one hand, we feed both $\mathcal{D}_{\mathrm{l}}$ and $\mathcal{D}_{\mathrm{ul}}$ into AdvMIL for \textit{$k$-fold semi-supervised} training (refer to Algorithm \ref{apx:alg1}), and test the performance of semi-supervised AdvMIL models on $\mathcal{D}_{\mathrm{ul}}$ and $\mathcal{D}_{\mathrm{test}}$. (II) On the other hand, we only feed $\mathcal{D}_{\mathrm{l}}$ into AdvMIL for \textit{fully-supervised} training, as the baseline for comparisons. We also test the performance of fully-supervised AdvMIL models on $\mathcal{D}_{\mathrm{ul}}$ and $\mathcal{D}_{\mathrm{test}}$. Except training data, other settings keep the same. The network ESAT + AdvMIL is adopted in this experiment, since it can achieve the best overall performance across three datasets (see Table \ref{tb1}).

\begin{figure}[h]
\centering
\includegraphics[width=0.43\textwidth]{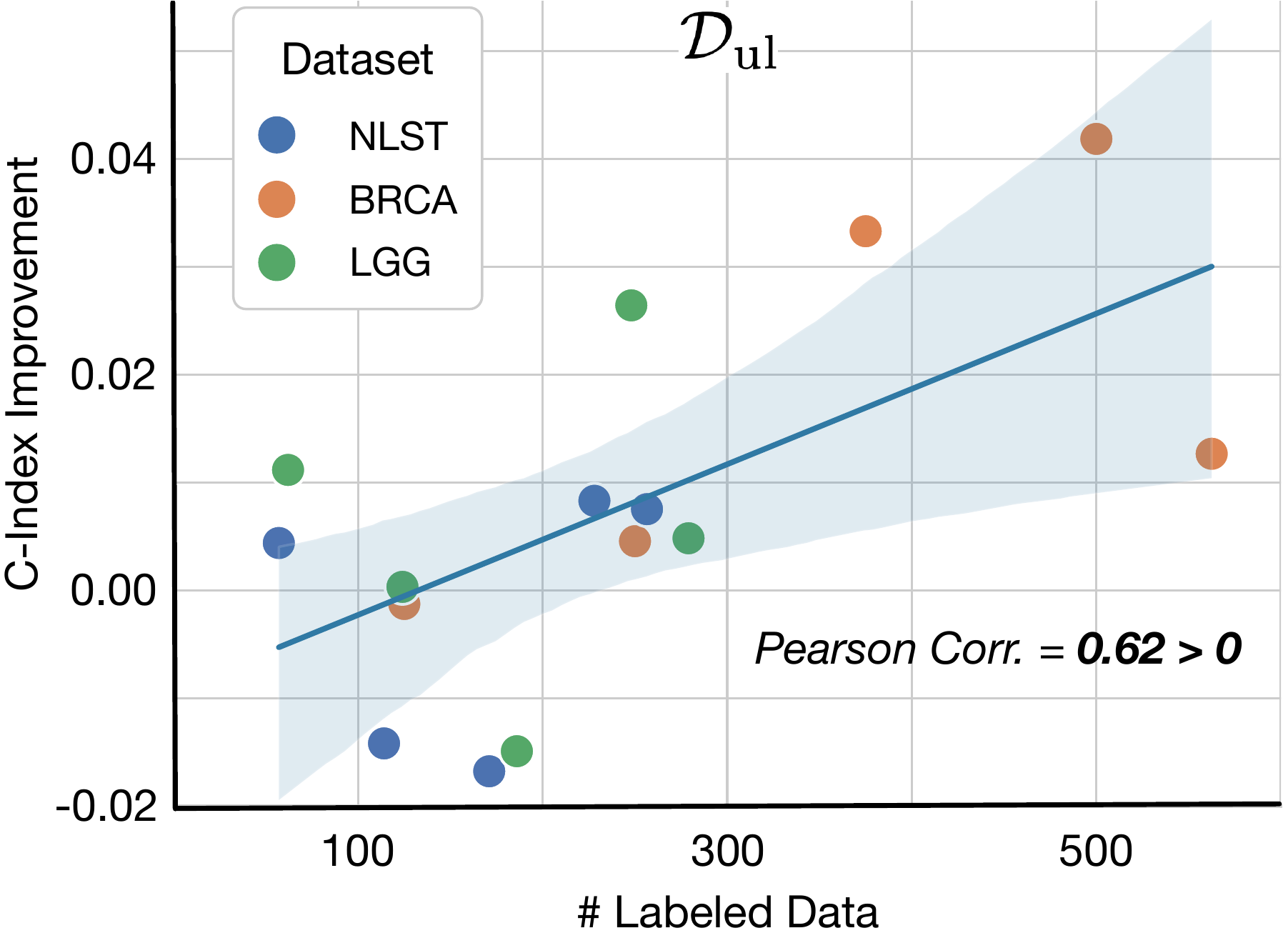}
\caption{Performance improvement of $k$-fold semi-supervised models over fully-supervised ones on $\mathcal{D}_{\mathrm{ul}}$. A linear regression of all points is plotted, with a 95\% confidence region. ``Corr.'' means ``Correlation''.}
\label{fig2a}
\end{figure}

\begin{figure}[h]
\centering
\includegraphics[width=0.43\textwidth]{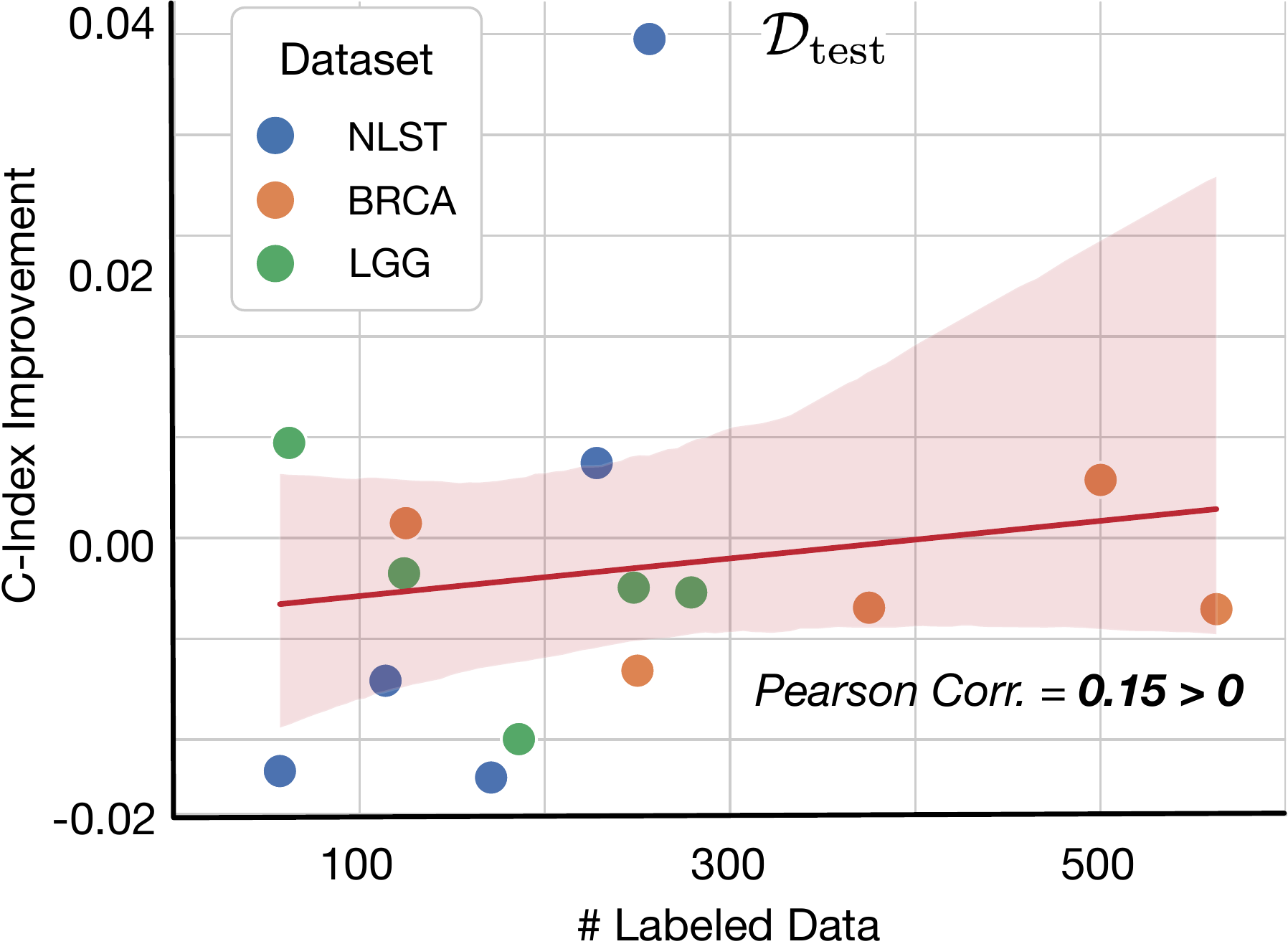}
\caption{Performance improvement of $k$-fold semi-supervised models over fully-supervised ones on $\mathcal{D}_{\mathrm{test}}$.}
\label{fig2b}
\end{figure}

We calculate the performance improvement of semi-supervised AdvMIL models over fully-supervised ones. Then we visualize this improvement on the vertical axis of 2D plane, along with the size of $\mathcal{D}_{\mathrm{l}}$ on the horizontal axis. A total of 15 points are obtained. The results on $\mathcal{D}_{\mathrm{ul}}$ and $\mathcal{D}_{\mathrm{test}}$ are shown in Figure \ref{fig2a} and Figure \ref{fig2b}, respectively. Full numerical results can be found in Table \ref{apx:tb5}. 

From these results, we can see that 1) on $\mathcal{D}_{\mathrm{ul}}$, the improvements of C-Index are often positive (10 out of 15 points) and they tend to become greater with increasing the size of $\mathcal{D}_{\mathrm{l}}$ (Pearson Corr. = 0.62); 2) on $\mathcal{D}_{\mathrm{test}}$, the improvements of C-Index are positive in a few cases (5 out of 15 points), and meanwhile they have a weakly-positive correlation with the size of $\mathcal{D}_{\mathrm{l}}$ (Pearson Corr. = 0.15). These findings mean that AdvMIL could be effective for predicting the unlabeled data \textit{used} in training, while often losing its effectiveness for predicting those \textit{unused} in training. Namely, AdvMIL could effectively utilize unlabeled WSIs in semi-supervised learning and often obtain superior performance on these unlabeled data; whereas the previous studies on WSI survival analysis have not paid enough attention on this. Moreover, increasing the size of labeled training data set would help to enlarge the C-Index improvement on unlabeled data $\mathcal{D}_{\mathrm{ul}}$. 

In addition, from Table \ref{apx:tb5}, we empirically find that our $k$-fold strategy ($k>1$) could often lead to better results than the traditional one without fold splitting ($i.e, k=1$), especially when the number of labeled data is small, namely when $\frac{\mathcal{D}_{\mathrm{l}}}{\mathcal{D}_{\mathrm{train}}}$ is 0.2, 0.4 or 0.6. These results confirm the effectiveness of our $k$-fold semi-supervised training strategy. 

\begin{table*}[tp]
\small
\centering
\caption{$k$-fold semi-supervised training with AdvMIL. Only using $\mathcal{D}_{\mathrm{l}}$ as training data indicates fully-supervised training. Using both $\mathcal{D}_{\mathrm{l}}$ and $\mathcal{D}_{\mathrm{ul}}$ as training data indicates semi-supervised training. $k=1$ means the traditional semi-supervised training without the fold splitting on $\mathcal{D}_{\mathrm{ul}}$. We train ESAT + AdvMIL on training data and measure its C-Index performances on $\mathcal{D}_{\mathrm{ul}}$ and $\mathcal{D}_{\mathrm{test}}$.}
\label{apx:tb5}
\begin{tabular}{cccccccc}
\toprule
\multirow{2}{*}{$\mathcal{D}_{\mathrm{l}}\ /\ \mathcal{D}_{\mathrm{train}}$} & \multirow{2}{*}{Training data} & \multicolumn{3}{c}{$\mathcal{D}_{\mathrm{ul}}$ (C-Index)}               & \multicolumn{3}{c}{$\mathcal{D}_{\mathrm{test}}$ (C-Index)}                   \\ \cmidrule(lr){3-5} \cmidrule(lr){6-8}
                                &                                & NLST        & BRCA        & LGG         & NLST          & BRCA          & LGG           \\ \midrule
\multirow{4}{*}{0.2}         & $\mathcal{D}_{\mathrm{l}}$                          & $0.616_{0.047}$ & $\textbf{0.553}_{0.029}$ & $0.619_{0.034}$ & $\textbf{0.623}_{0.082}$ & $0.530_{0.062}$ & $0.602_{0.055}$ \\
                                & $\mathcal{D}_{\mathrm{l}}$ + $\mathcal{D}_{\mathrm{ul}}$ ($k$=1)             & $0.578_{0.053}$ & $0.545_{0.043}$ & $0.622_{0.022}$ & $0.576_{0.055}$ & $\textbf{0.540}_{0.093}$ & $0.608_{0.041}$ \\
                                & $\mathcal{D}_{\mathrm{l}}$ + $\mathcal{D}_{\mathrm{ul}}$ ($k$=3)             & $0.597_{0.049}$ & $0.542_{0.047}$ & $0.621_{0.031}$ & $0.616_{0.077}$ & $0.518_{0.097}$ & $0.592_{0.059}$ \\
                                & $\mathcal{D}_{\mathrm{l}}$ + $\mathcal{D}_{\mathrm{ul}}$ ($k$=5)             & $\textbf{0.620}_{0.047}$ & $0.552_{0.049}$ & $\textbf{0.630}_{0.019}$ & $0.610_{0.065}$ & $0.528_{0.080}$ & $\textbf{0.610}_{0.048}$ \\ \midrule
\multirow{4}{*}{0.4}            & $\mathcal{D}_{\mathrm{l}}$ & $\textbf{0.632}_{0.047}$ & $0.574_{0.042}$ & $0.603_{0.053}$ & $\textbf{0.622}_{0.060}$ & $\textbf{0.579}_{0.135}$ & $\textbf{0.597}_{0.066}$ \\
                                & $\mathcal{D}_{\mathrm{l}}$ + $\mathcal{D}_{\mathrm{ul}}$ ($k$=1)             & $0.602_{0.027}$ & $0.524_{0.052}$ & $0.595_{0.090}$ & $0.585_{0.065}$ & $0.527_{0.085}$ & $0.580_{0.099}$ \\
                                & $\mathcal{D}_{\mathrm{l}}$ + $\mathcal{D}_{\mathrm{ul}}$ ($k$=3)            & $0.618_{0.051}$ & $0.549_{0.080}$ & $\textbf{0.604}_{0.062}$ & $0.603_{0.054}$ & $0.536_{0.069}$ & $\textbf{0.597}_{0.076}$ \\
                                & $\mathcal{D}_{\mathrm{l}}$ + $\mathcal{D}_{\mathrm{ul}}$ ($k$=5)            & $0.606_{0.050}$ & $\textbf{0.578}_{0.046}$ & $0.592_{0.085}$ & $0.586_{0.077}$ & $0.576_{0.037}$ & $0.595_{0.079}$ \\ \midrule
\multirow{4}{*}{0.6}            & $\mathcal{D}_{\mathrm{l}}$ & $\textbf{0.637}_{0.054}$ & $0.532_{0.058}$ & $\textbf{0.623}_{0.046}$ & $\textbf{0.646}_{0.043}$ & $0.565_{0.097}$ & $\textbf{0.649}_{0.045}$ \\
                                & $\mathcal{D}_{\mathrm{l}}$ + $\mathcal{D}_{\mathrm{ul}}$ ($k$=1)             & $0.612_{0.032}$ & $0.556_{0.030}$ & $0.606_{0.027}$ & $0.610_{0.083}$ & $0.563_{0.092}$ & $0.639_{0.045}$ \\
                                & $\mathcal{D}_{\mathrm{l}}$ + $\mathcal{D}_{\mathrm{ul}}$ ($k$=3)             & $0.618_{0.059}$ & $0.535_{0.060}$ & $0.601_{0.055}$ & $0.645_{0.040}$ & $0.533_{0.063}$ & $0.637_{0.063}$ \\
                                & $\mathcal{D}_{\mathrm{l}}$ + $\mathcal{D}_{\mathrm{ul}}$ ($k$=5)             & $0.620_{0.057}$ & $\textbf{0.566}_{0.060}$ & $0.608_{0.031}$ & $0.632_{0.059}$ & $\textbf{0.569}_{0.097}$ & $0.625_{0.046}$ \\ \midrule
\multirow{4}{*}{0.8}            & $\mathcal{D}_{\mathrm{l}}$ & $0.629_{0.044}$ & $0.517_{0.062}$ & $0.609_{0.048}$ & $0.628_{0.062}$ & $0.538_{0.089}$ & $0.630_{0.082}$ \\
                                & $\mathcal{D}_{\mathrm{l}}$ + $\mathcal{D}_{\mathrm{ul}}$ ($k$=1)             & $\textbf{0.638}_{0.049}$ & $0.524_{0.065}$ & $0.590_{0.039}$ & $0.643_{0.048}$ & $0.552_{0.072}$ & $0.614_{0.080}$ \\
                                & $\mathcal{D}_{\mathrm{l}}$ + $\mathcal{D}_{\mathrm{ul}}$ ($k$=3)            & $0.623_{0.075}$ & $0.511_{0.086}$ & $\textbf{0.635}_{0.092}$ & $0.642_{0.048}$ & $\textbf{0.567}_{0.093}$ & $0.630_{0.085}$  \\
                                & $\mathcal{D}_{\mathrm{l}}$ + $\mathcal{D}_{\mathrm{ul}}$ ($k$=5)             & $0.622_{0.032}$ & $\textbf{0.559}_{0.069}$ & $0.615_{0.038}$ & $\textbf{0.646}_{0.046}$ & $0.554_{0.085}$ & $\textbf{0.634}_{0.093}$ \\ \midrule
\multirow{4}{*}{0.9}            & $\mathcal{D}_{\mathrm{l}}$ & $0.713_{0.047}$ & $0.470_{0.139}$ & $0.552_{0.134}$ & $0.661_{0.038}$ & $0.556_{0.055}$ & $0.637_{0.085}$ \\
                                & $\mathcal{D}_{\mathrm{l}}$ + $\mathcal{D}_{\mathrm{ul}}$ ($k$=1)            & $0.716_{0.037}$ & $\textbf{0.482}_{0.187}$ & $0.550_{0.115}$ & $0.650_{0.062}$ & $\textbf{0.559}_{0.070}$ & $\textbf{0.641}_{0.066}$ \\
                                & $\mathcal{D}_{\mathrm{l}}$ + $\mathcal{D}_{\mathrm{ul}}$ ($k$=3)             & $\textbf{0.720}_{0.069}$ & $0.449_{0.202}$ & $\textbf{0.557}_{0.117}$ & $0.654_{0.055}$ & $0.555_{0.095}$ & $0.619_{0.050}$ \\
                                & $\mathcal{D}_{\mathrm{l}}$ + $\mathcal{D}_{\mathrm{ul}}$ ($k$=5)             & $0.699_{0.048}$ & $0.380_{0.154}$ & $0.545_{0.104}$ & $\textbf{0.664}_{0.051}$ & $0.521_{0.074}$ & $0.637_{0.060}$ \\ \bottomrule
\end{tabular}
\end{table*}

\begin{table*}[tp]
\small
\centering
\caption{Ablation study on region-level instance projection (RLIP). $\Delta$ denotes the improvement of RLIP over regular projection. ESAT + AdvMIL is used.}
\label{tb2}
\begin{tabular}{cccccccc}
\toprule
\multirow{2}{*}{Noise} & \multirow{2}{*}{Fusion operation} & \multicolumn{3}{c}{C-Index $\uparrow$}             & \multicolumn{3}{c}{MAE $\downarrow$}                 \\ \cmidrule(lr){3-5} \cmidrule(lr){6-8}
                       &            & NLST        & BRCA        & LGG         & NLST        & BRCA        & LGG         \\ \midrule
\multirow{3}{*}{0-1} & Projection             & $0.658_{0.047}$ & $0.543_{0.072}$ & $\textbf{0.624}_{0.084}$ & $0.1983_{0.0228}$ & $\textbf{0.0375}_{0.0082}$ & $\textbf{0.0519}_{0.0064}$ \\
                     & RLIP                  & $\textbf{0.672}_{0.048}$ & $\textbf{0.545}_{0.065}$ & $0.621_{0.063}$ & $\textbf{0.1871}_{0.0203}$ & $0.0383_{0.0081}$ & $0.0526_{0.0058}$ \\ \cmidrule(lr){2-2} 
                     & $\Delta$  & + 2.13\% & + 0.37\% & - 0.48\% & - 5.65\% & + 2.13\% & + 1.35\% \\ \midrule
\multirow{3}{*}{1-0} & Projection                          & $0.616_{0.052}$ & $0.548_{0.084}$ & $0.641_{0.079}$ & $0.2015_{0.0243}$ & $\textbf{0.0344}_{0.0063}$ & $\textbf{0.0518}_{0.0053}$ \\
                     & RLIP                  & $\textbf{0.649}_{0.039}$ & $\textbf{0.562}_{0.067}$ & $\textbf{0.642}_{0.076}$ & $\textbf{0.1995}_{0.0240}$ & $0.0349_{0.0065}$ & $0.0522_{0.0049}$ \\ \cmidrule(lr){2-2}  
                     & $\Delta$ & + 5.36\% & + 2.55\% & + 0.16\% & - 9.93\% & + 1.45\% & + 0.77\% \\ \midrule
\multirow{3}{*}{1-1} & Projection                           & $0.652_{0.032}$ & $0.531_{0.080}$ & $0.616_{0.082}$ & $0.1996_{0.0213}$ & $0.0389_{0.0054}$ & $0.0537_{0.0049}$ \\
                     & RLIP                   & $\textbf{0.660}_{0.042}$ & $\textbf{0.545}_{0.075}$ & $\textbf{0.634}_{0.086}$ & $\textbf{0.1849}_{0.0153}$ & $\textbf{0.0366}_{0.0065}$ & $\textbf{0.0523}_{0.0051}$ \\ \cmidrule(lr){2-2}
                     & $\Delta$ & + 1.23\% & + 2.64\% & + 2.92\% & - 7.36\% & - 5.91\% & - 2.61\% \\ \bottomrule
\end{tabular}
\end{table*}

\begin{table*}[tp]
\small
\centering
\caption{Noise effect on the performance of ESAT + AdvMIL. The noise with \emph{Uniform} distribution could often perform better than that with \emph{Gaussian} distribution.}
\label{apx:tb3}
\begin{tabular}{cccccccc}
\toprule
\multicolumn{2}{c}{\multirow{2}{*}{$\mathcal{N}$}} & \multicolumn{3}{c}{C-Index $\uparrow$}             & \multicolumn{3}{c}{MAE $\downarrow$}                 \\ \cmidrule(lr){3-5} \cmidrule(lr){6-8}
                    &               & NLST        & BRCA        & LGG         & NLST        & BRCA        & LGG         \\ \midrule
\multirow{3}{*}{Uniform(0,1)} & 0-1                   & $\textbf{0.672}_{0.048}$ & $0.545_{0.065}$ & $0.621_{0.063}$ & $0.1871_{0.0203}$ & $0.0383_{0.0081}$ & $0.0526_{0.0058}$ \\
& 1-0                   & $0.649_{0.039}$ & $\textbf{0.562}_{0.067}$ & $\textbf{0.642}_{0.076}$ & $0.1995_{0.0240}$ & $\textbf{0.0349}_{0.0065}$ & $\textbf{0.0522}_{0.0049}$ \\ 
& 1-1                   & $0.660_{0.042}$ & $0.545_{0.075}$ & $0.634_{0.086}$ & $\textbf{0.1849}_{0.0153}$ & $0.0366_{0.0065}$ & $0.0523_{0.0051}$ \\ \cmidrule(lr){1-2}
\multirow{3}{*}{Gaussian(0,1)} & 0-1                   & $0.632_{0.039}$ & $0.508_{0.102}$ & $0.631_{0.090}$ & $0.1928_{0.0211}$ & $0.0392_{0.0063}$ & $0.0567_{0.0080}$ \\
& 1-0                   & $\textbf{0.634}_{0.042}$ & $0.461_{0.078}$ & $0.624_{0.050}$ & $0.2011_{0.0264}$ & $\textbf{0.0363}_{0.0060}$ & $\textbf{0.0529}_{0.0033}$ \\ 
& 1-1                   & $0.590_{0.072}$ & $\textbf{0.554}_{0.047}$ & $\textbf{0.646}_{0.051}$ & $\textbf{0.2019}_{0.0223}$ & $0.0403_{0.0078}$ & $0.0556_{0.0071}$ \\  \bottomrule
\end{tabular}
\end{table*}

There are many studies that demonstrate GAN to be a promising and effective means for semi-supervised learning \citep{springenberg2015unsupervised,salimans2016improved,miyato2018virtual,li2021triple,gui2021areview}. Our experiments also verify this, even when it is a small data regime (only around 500 patients in our task) in the field of WSI survival analysis. 

\begin{figure*}[tp]
\centering
\includegraphics[width=1.00\textwidth]{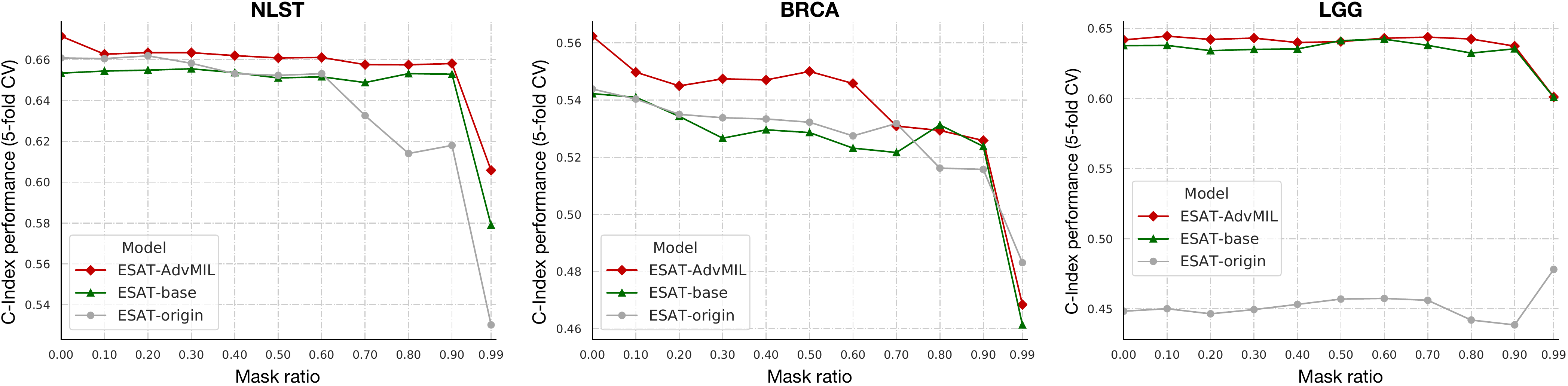}
\caption{Robustness against patch occlusion. A mask ratio of 0.99 indicates that there is only one region (with a size of $4,096\times 4,096$ pixels at the highest magnification) reserved in each test WSI.}
\label{apx:fig-rst-occ}
\end{figure*}

\subsection{Ablation study and analysis}
(1) \emph{Study on region-level instance projection}

We validate the region-level instance projection (RLIP) strategy that is proposed for the feature fusion in $D$, and compare it with a regular strategy, WSI-level projection, named \textit{Projection}. This regular strategy directly projects a conditional vector on the global feature of WSI. It skips the smooth transition (used in RLIP) from patch to region and region to WSI. Note that the \textit{Projection}, which makes a simple and direct fusion on WSI and survival time, is a strategy having not appeared in previous models. It is just used as a baseline for comparisons. 

From the results in Table \ref{tb2}, we can summarize that 1) RLIP almost always gains C-Index rises (the highest one is + 5.36\%) on three datasets; 2) in most cases RLIP could decrease MAEs by large margins (2.61\% -- 9.93\%), and only in a few cases, the MAEs of RLIP increase by no larger than 2.13\%; 3) when setting noise $\mathcal{N}$ to 1-1, RLIP can completely surpass \textit{Projection} on three datasets in terms of both C-Index and MAE. These empirical results demonstrate the effectiveness of our region-level strategy, and suggest that an early feature fusion, \textit{i.e.}, region-level projection rather than a direct WSI-level projection, may be more likely to better incorporate conditional information and help adversarial multiple instance learning. 

(2) \emph{Study on noise types}
\label{sec:ablation}

We further test the effect of noise type on model performance. Two widely-used distributions, \textit{Uniform} and \textit{Gaussian}, are tested. From the results shown in Table \ref{apx:tb3}, we can see that Uniform distribution could often obtain better performances. Especially in terms of MAE, Uniform ones always have obvious advantages over Gaussian ones. One possible reason is that when sampling $\mathcal{N}$ from Gaussian distribution, the estimation of time-to-event distribution via implicitly sampling would be more concentrated and it thereby would tend to enlarge the bias of predictions. 

In addition, as mentioned in Section \ref{sec:overperf}, we argue that the noise, which is combined together with conditional input for prediction, may be a critical factor affecting model performance. Some techniques like adversarial training \citep{goodfellow2014explaining,goodfellow2016nips} may be a promising means to validate this argument. We leave this in the future work, since adversarial training is yet another research topic and it is not the focus of this paper. 

\subsection{Robustness analysis}
We carry out experiments on three used WSI datasets to validate the robustness of AdvMIL-based models. Next, we show how we set up experiments for robustness analysis, and then present experimental results and findings. 

Specifically, we evaluate the model’s robustness to patch-level transformations, including image occlusion, Gaussian blurring and HED (Hematoxylin-Eosin-DAB) color space perturbation, in view of the following observations. (I) It is really challenging to assess model robustness by directly transforming an entire gigapixel WSI into various views, unlike the straightforward operations on most natural images with common sizes. (II) Patch occlusion has been utilized as an experimental scenario by previous works \citep{Naseer2021,Chen2022transmix} to study the robustness of prevalent computer vision models. (III) Gaussian blurring is a classical transformation for natural images to simulate out-of-focus artifacts, also widely utilized for histological WSIs \citep{Tellez2019,Cheng2021robust}. (IV) HED color perturbation is specifically designed for H\&E images \citep{Tellez2018} to simulate a color variation routine, which is frequently adopted in computational pathology \citep{Tellez2019} and could be more suitable for our applications. 

(1) \emph{Robustness to patch occlusion}

We randomly mask the patches of each WSI (in test sets), according to various mask ratios. A mask ratio of 0.99 indicates that there is only one region (with a size of $4,096\times 4,096$ pixels at the highest magnification) reserved in each test WSI. The models, which have been fitted on complete training data previously, are adopted for test. The C-Index averaged on 5 folds is reported and shown in Figure \ref{apx:fig-rst-occ}. 

From Figure \ref{apx:fig-rst-occ}, we can observe that AdvMIL-based models almost always outperform two baselines on various occlusion levels across three datasets. Moreover, it is found that ESAT + AdvMIL has more stable C-Index performance than both ESAT$_{\mathrm{origin}}$ and ESAT$_{\mathrm{base}}$ on NLST and LGG, when mask ratio is larger than 0.5. These experimental results demonstrate that our adversarial survival analysis models could often be more robust against occlusion than traditional ones on WSIs. In other words, the occlusion robustness of traditional WSI survival analysis models could be further enhanced via our adversarial MIL scheme. 

Besides, we interestingly see that some meaningful results are still achieved by AdvMIL-based models (C-Index $\approx$ 0.6) on NLST and LGG, even when mask ratio is 0.99, \textit{i.e.}, only reserving one region in each test WSI. The underlying reasons behind this phenomenon could have three-fold: i) the information abundance and sparsity inherent in histological WSIs \citep{zarella2018apractical}; ii) the intrinsic robustness of Transformer-based models, demonstrated in \cite{Naseer2021} and \cite{Ghaffari2022adv}; iii) the robustness further improved by our AdvMIL, seen from the improvement of AdvMIL-based ESAT over its non-adversarial counterpart ESAT$_{\mathrm{base}}$. 

(2) \emph{Robustness to image noises}

\begin{figure}[tp]
\centering
\includegraphics[width=0.48\textwidth]{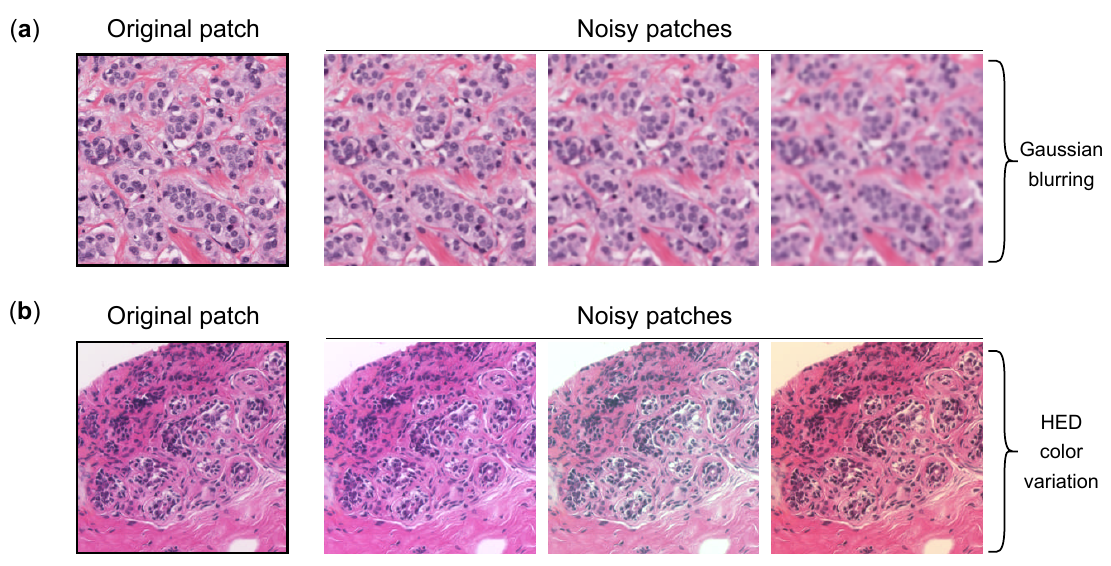}
\caption{Examples of (a) Gaussian blurring and (b) HED color variation.}
\label{apx:fig-noisy-imgs}
\end{figure}

We apply two representative noises to patch images, \textit{i.e.}, Gaussian blurring (Gau. blur) and HED color variation (HED var.). Some examples of them are shown in Figure \ref{apx:fig-noisy-imgs}. Similarly, those trained models are adopted for testing on noisy patch images. We choose the first fold in each dataset for experiments, as there is a huge total number of patch images (nearly 10,000,000 images of 256 $\times$ 256 pixels) to be processed in the 3,101 gigapixel WSIs used by us. 

From the experimental results shown in Table \ref{apx:tb-rst-noise}, we clearly see that AdvMIL-based models always obtain the best C-Index across three datasets, after applying image noises. Moreover, we also see a more stable performance on AdvMIL-based models in this experiment. For example, the biggest changes in C-Index of ESAT + AdvMIL are 0.004, 0.03, and 0.02 on NLST, BRCA, and LGG, respectively; while those of ESAT$_{\mathrm{base}}$ are 0.012, 0.055, and 0.028, apparently larger than those of AdvMIL-based models. ESAT$_{\mathrm{origin}}$ also shows similar results, except its nonsense results (C-Index $\leq$ 0.5) on LGG. And most notably, on BRCA, the C-Index of ESAT$_{\mathrm{base}}$ decreases by 0.055 and 0.044 after applying Gaussian blurring and HED color variation, respectively; while that of ESAT + AdvMIL still keeps its superiority without any decline. 

These empirical observations indicate the potential advantage of adversarial MIL models in robustness over traditional ones. One explicit and intuitive explanation for this is the mechanism of adversarial learning, \textit{i.e.}, adversarial loss could help to optimize generator, and make generator become more robust to the input space extended and smoothed by noises, leading to better performance than non-adversarial ones in noisy environments, as stated in \cite{miyato2018virtual}. 

\begin{table}[tp]
\small
\centering
\caption{Robustness against patch image noises. C-Index is reported on the test set of the first fold.}
\label{apx:tb-rst-noise}
\begin{tabular}{ccp{1.1cm}<{\centering}cc}
\toprule
\multirow{2}{*}{Dataset} & \multirow{2}{*}{Model} & \multicolumn{3}{c}{Image noise} \\ \cmidrule(lr){3-5}
                         &                        & ---     & Gau. blur    & HED var.    \\ \midrule
\multirow{3}{*}{NLST}    & ESAT$_{\text{origin}}$                 & 0.642    &  0.661   &  0.667       \\
                         & ESAT$_{\text{base}}$                   & 0.663    &  0.672   &  0.675       \\
                         & ESAT + AdvMIL                 & \textbf{0.682}    &  \textbf{0.680}   & \textbf{0.684}   \\ \midrule
\multirow{3}{*}{BRCA}    & ESAT$_{\text{origin}}$                 & 0.543    &  0.589   &  0.588       \\
                         & ESAT$_{\text{base}}$                   & 0.601    &  0.546   &  0.557       \\
                         & ESAT + AdvMIL                 & \textbf{0.602}    &  \textbf{0.620}   & \textbf{0.632}   \\ \midrule
\multirow{3}{*}{LGG}    & ESAT$_{\text{origin}}$                  & 0.468    &  0.492   &  0.498       \\
                         & ESAT$_{\text{base}}$                   & 0.554    &  0.579   &  0.582       \\
                         & ESAT + AdvMIL                 & \textbf{0.604}    &  \textbf{0.584}   & \textbf{0.597}   \\ \bottomrule
\end{tabular}
\end{table}

\begin{figure*}[tp]
\centering
\includegraphics[width=0.96\textwidth]{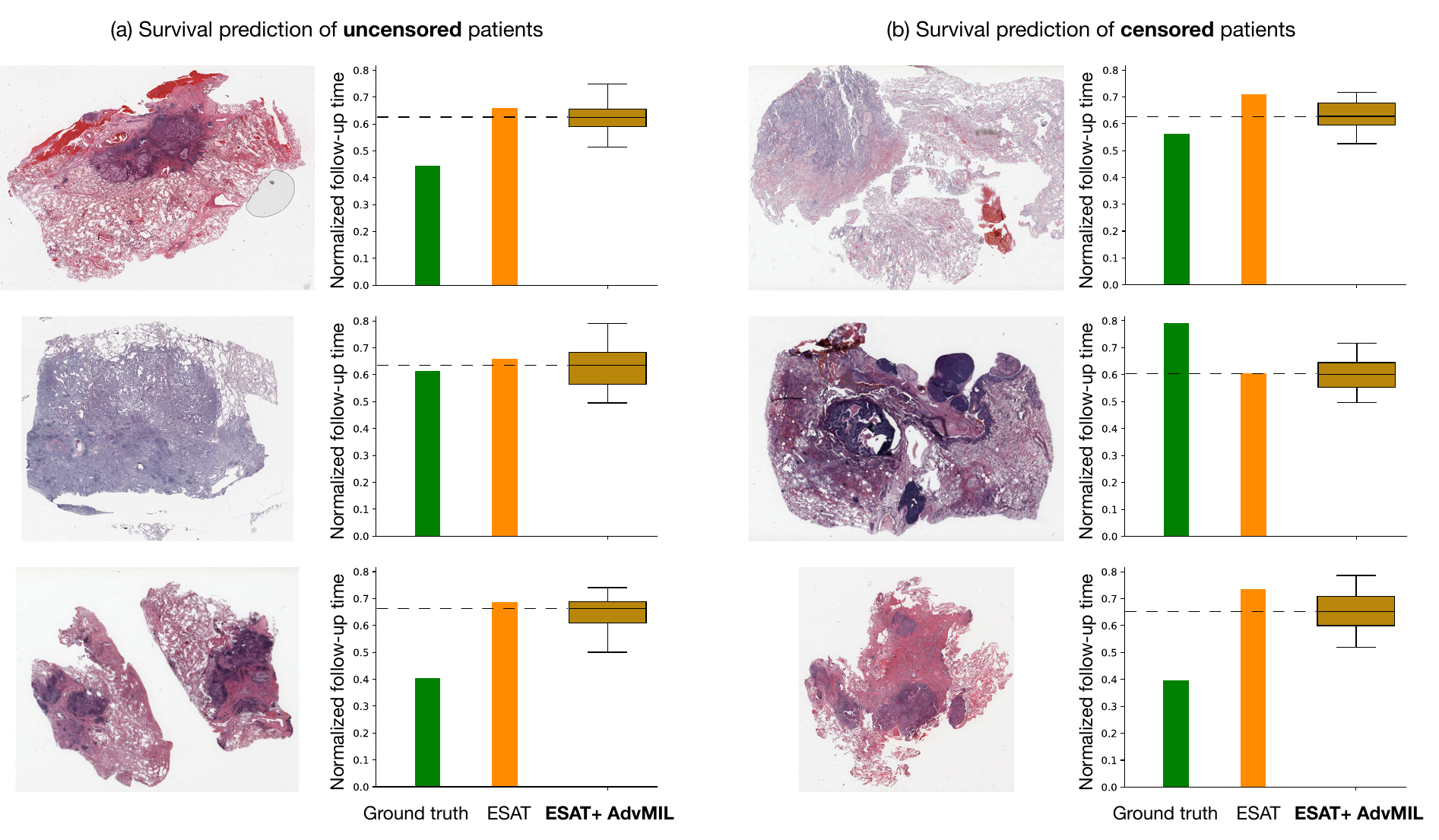}
\caption{Case study of the time-to-event estimations given by ESAT (a base version) and its AdvMIL-improved version. Three uncensored patients (left) and three censored patients (right) are randomly selected from the test set of NLST as study cases.}
\label{apx:fig0}
\end{figure*}

\subsection{Case study and analysis}
We randomly select 3 uncensored patients ($\delta=0$, with event occurrence) and 3 censored patients ($\delta=1$, without event occurrence) from the test set of NLST. We show their time-to-event estimations by ESAT$_\mathrm{base}$ and ESAT + AdvMIL, as well as their last follow-up times, in Figure \ref{apx:fig0}. 

From these results, we can see that 1) AdvMIL enables ESAT to provide many time-to-event estimations for one patient, while original ESAT only gives one single completely-certain result; 2) the distribution estimation given by AdvMIL often covers the single point estimation given by ESAT; 3) the median of distribution estimation is often closer to the ground truth than point estimation. However, we cannot quantitatively assess the goodness of distribution coverage, because the ground truth of time-to-event distribution is unknown. 

\section{Discussion} 
Histological WSIs contain rich microenvironmental cues that are vital for disease prognosis. An accurate assessment of WSI-based prognosis could help to improve patient management and disease outcomes. Although many end-to-end weakly-supervised models have been developed to estimate patient prognosis from WSIs, their potential is generally restricted by the classical paradigm of survival analysis and fully-supervised learning. Inspired by adversarial time-to-event modeling, we propose a novel adversarial MIL framework to exploit their potential. This framework could bring new vigor and vitality into 
the survival analysis in computational pathology, by enabling existing MIL networks to estimate a more robust time-to-event distribution and learn from unlabeled WSI data via semi-supervised training, at a relatively low computational cost. 

We emphasize that this study stands on the shoulders of GANs \citep{goodfellow2014generative,mirza2014conditional} and DATE \citep{chapfuwa2018adversarial}, and shows for the first time how to generalize them to the MIL that is much necessary for WSI representation learning. Last but not least, apart from the technical contributions made by AdvMIL, this study also would like to highlight its practical contribution (\textit{e.g.}, semi-supervised applications) to computational pathology community. 

The main limitations of this study include that the coverage goodness of distribution estimation cannot be quantitatively evaluated since the ground truth of patient survival distribution is unavailable. In addition, there are also some constraints in our experiments: 1) limited sampling times are adopted to return distribution estimation because it is relatively time-consuming to infer once on gigapixel images, and 2) our datasets are limited to three cancer types, not covering more variety and clinical comparisons. 

\section{Conclusions}
This paper proposes a novel framework, adversarial multiple instance learning (AdvMIL), for the survival analysis on gigapixel WSIs. This framework generalizes adversarial time-to-event modeling to MIL mainly by its two cores: the MIL encoder in generator and the fusion network with region-level instance projection in discriminator. It is a plug-and-play framework; namely it can be easily and efficiently applied to most end-to-end MIL models. The extensive experiments on three WSI datasets demonstrate that the proposed AdvMIL framework could often bring performance improvement to existing mainstream MIL models at a relatively low computational cost. Most importantly, AdvMIL could assist these existing models to fulfill a more effective estimation of time-to-event distribution and effectively learn from unlabeled WSIs via semi-supervised learning. Moreover, it is also observed that AdvMIL could help improving the robustness of models against patch occlusion and two representative image noises.

\bibliographystyle{model2-names.bst}\biboptions{authoryear}
\bibliography{ref-advmil}

\end{document}